\documentclass[11pt]{article}
\usepackage{amsfonts,latexsym,graphicx,ifpdf,overpic,subfigure,url,wrapfig,enumerate}
\usepackage{amsmath}
\usepackage{amsthm}
\advance \topmargin by -\headheight
\advance \topmargin by -\headsep
\evensidemargin \oddsidemargin
\usepackage
  [breaklinks,bookmarks,bookmarksnumbered,bookmarksopen,bookmarksopenlevel=2]
  {hyperref}
{\makeatletter \hypersetup{pdftitle={\@title}}}

\ifpdf
  \usepackage{epstopdf}
  \makeatletter
  \edef\Gin@extensions{.eps,\Gin@extensions}
  \makeatother
\fi

\ifpdf
  \let\enablecleaneps=\relax
  
\else
  \def\enablecleaneps{\DeclareGraphicsRule{.eps}{eps}{.eps}{`eps2eps ##1 -}}
  
  \enablecleaneps
\fi

\let\latexcite=\cite
\def\cite{\nolinebreak\latexcite}
\let\latexref=\ref
\def\ref{\nolinebreak\latexref}

\newtheorem{theorem}{Theorem}
\newtheorem{proposition}{Proposition}
\newtheorem{lemma}[theorem]{Lemma}
\newtheorem{corollary}[theorem]{Corollary}
\newtheorem{onerule}{Rule}

\let\epsilon=\varepsilon

\begin{document}

\title{Locked and Unlocked Chains of Planar Shapes%
  \thanks{%
A preliminary extendend abstract of this paper appeared in
\emph{Proceedings of the 22nd Annual ACM Symposium on Computational Geometry}, 
June 2006, \protect \cite{conf-version}.
R. Connelly is supported in part by NSF grant DMS-0209595.
E.~Demaine is supported in part by NSF grant CCF-0347776
      and DOE grant DE-FG02-04ER25647.
S.~Langerman is Chercheur qualifi\'e du FNRS.
J.~Mitchell is supported in part by
      NSF
      (CCF-0431030, CCF-0528209, CCF-0729019),
      Metron Aviation, and NASA Ames.}}


\author{%
  Robert Connelly%
    \thanks{Department of Mathematics, Cornell University,
      Ithaca, NY 14853, USA.  \protect\url{connelly@math.cornell.edu}}
\and
  Erik D. Demaine%
    \thanks{MIT Computer Science and Artificial Intelligence Laboratory,
      32 Vassar St., Cambridge, MA 02139, USA,
      \protect\url{{edemaine,mdemaine}@mit.edu}}
\and
  Martin L. Demaine%
    \footnotemark[2]
\and
  S\'andor P. Fekete%
    \thanks{Department of Computer Science, Braunschweig University of Technology,
      M\"uhlenpfordtstr. 23, D-38106 Braunschweig, Germany.
      \protect\url{s.fekete@tu-bs.de}}
\and
  Stefan Langerman%
    \thanks{Chercheur qualifi\'e du FNRS, Universit\'e Libre de Bruxelles,
      D\'epartement d'informatique, ULB CP212, Belgium.
      \protect\url{Stefan.Langerman@ulb.ac.be}}
\and
  Joseph S. B. Mitchell%
    \thanks{Department of Applied Mathematics and Statistics,
      Stony Brook University, Stony Brook, NY 11794-3600, USA.
      \protect\url{jsbm@ams.sunysb.edu}}
\and
  Ares {Rib\'o}%
    \thanks{Institut f\"ur Informatik, Freie Universit\"at Berlin,
      Takustra{\ss}e 9, D-14195 Berlin, Germany.
      \protect\url{{ribo,rote}@inf.fu-berlin.de}}
\and
  G\"unter Rote%
    \footnotemark[6]
}

\date{}
\maketitle

\begin{abstract}
  We extend linkage unfolding results from the well-studied case of
  polygonal linkages to the more general case of linkages of polygons.
  More precisely, we consider chains of nonoverlapping rigid planar shapes
  (Jordan regions) that are hinged together sequentially at rotatable joints.
  Our goal is to characterize the familes of planar shapes that admit
  \emph{locked chains}, where some configurations cannot be reached by
  continuous reconfiguration without self-intersection, and which families of
  planar shapes guarantee \emph{universal foldability}, where every chain is
  guaranteed to have a connected configuration space.
  Previously, only obtuse triangles were known to admit locked shapes,
  and only line segments were known to guarantee universal foldability.
  We show that a surprisingly general family of planar shapes,
  called \emph{slender adornments}, guarantees universal foldability:
  roughly, the distance from each edge along the path along the boundary of the slender adornment to each hinge should be monotone.
  In contrast, we show that isosceles triangles with any desired apex angle
  $< 90^\circ$ admit locked chains, which is precisely the threshold beyond
  which the slender property no longer holds.
\end{abstract}



\section{Introduction}

In this paper, we explore the motion-planning problem of \emph{reconfiguring}
or \emph{folding} a hinged collection of rigid objects from one state to
another while avoiding self-intersection.
This general problem has been studied since the beginnings
of the motion-planning literature when Reif \cite{r-cmpg-79}
proved that deciding reconfigurability of a ``tree'' of polyhedra,
amidst fixed polyhedral obstacles, is PSPACE-hard.
This result has been strengthened in various directions over the years,
although the cleanest versions were obtained only very recently:
deciding reconfigurability of a tree of line segments in the plane, and
deciding reconfigurability of a chain of line segments in 3D,
are both PSPACE-complete \cite{Alt-Knauer-Rote-Whitesides-2004}.
This result is tight in the sense that deciding reconfigurability
of a chain of line segments in the plane is easy, in fact, trivial:
the answer is always yes \cite{Connelly-Demaine-Rote-2003}.

These results illustrate a rather fine line in reconfiguration problems between
computationally difficult and computationally trivial.  The goal of our work
is to characterize what families of planar shapes and hingings lead to the
latter outcome,
a \emph{universality result} that reconfiguration is always possible.
The only known example of such a result, however, is the family of chains
of line segments, and that problem was unsolved for about 25 years
\cite{Connelly-Demaine-Rote-2003}. (A \emph{chain} is a sequence of line segments joined end-to-end that are disjoint except except for consecutive endpoints, where they are hinged.  It is called an \emph{open chain} when the first line segment is not joined to the last, and \emph{closed} when it is joined to the last segment in a closed cycle.)   Even small perturbations to the problem,
such as allowing a single point where three line segments join, leads to
\emph{locked} examples where reconfiguration is impossible
\cite{Connelly-Demaine-Rote-2002-infinitesimally-locked}.

\begin{figure}[htbp]
  \centering
  \subfigure[]
    {\includegraphics[scale=0.7]{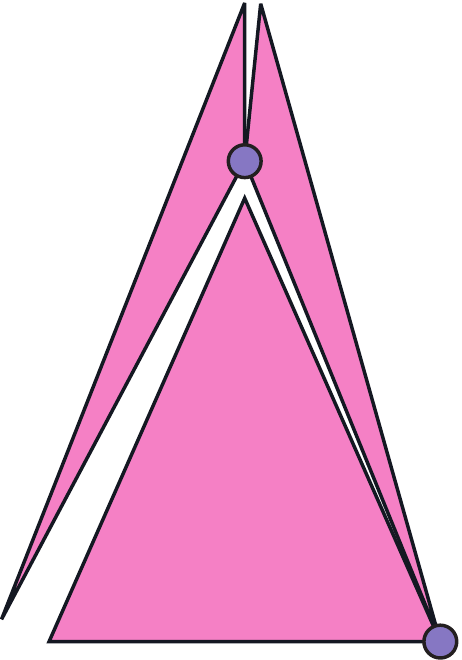}
     \label{locked 3 triangles}}\hfil\hfil
  \subfigure[]
    {\includegraphics[scale=0.7]{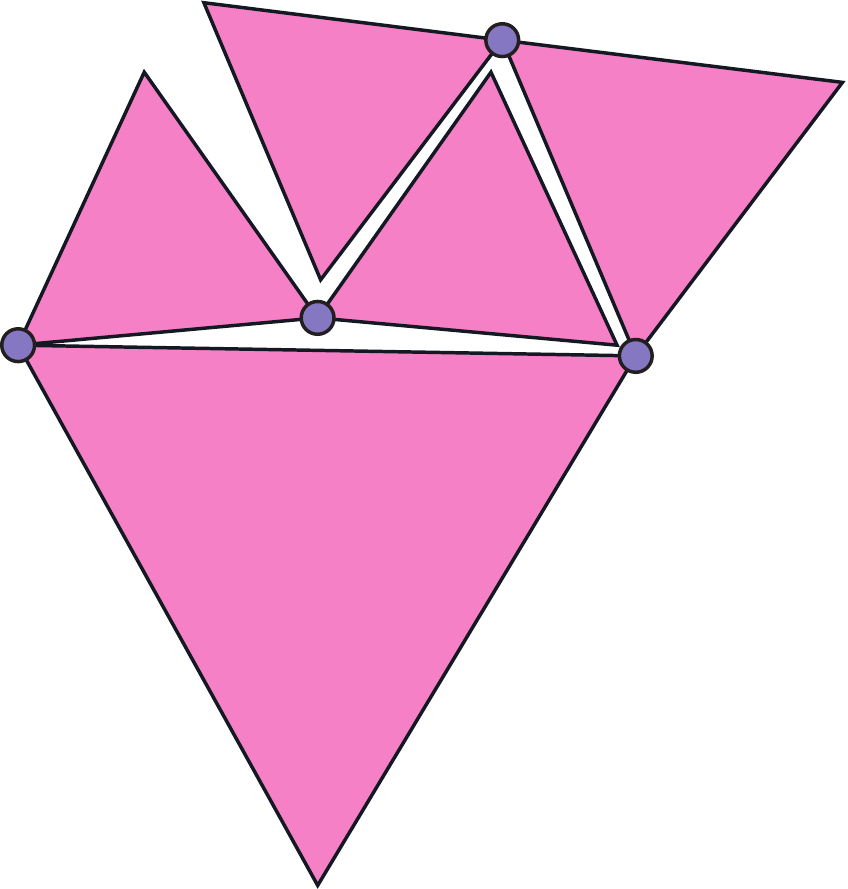}
     \label{locked unequal equilateral}}
  \caption{Simple examples of locked chains of triangles.
    (a) A locked chain of three triangles.
    (b) A locked chain of equilateral triangles of different sizes.
        The gaps should be tighter than drawn.}
  \label{locked few triangles}
\end{figure}

What about chains of shapes other than line segments?
It is easy to see that a shape tucked into a ``pocket'' of a nonconvex shape
immediately makes trivial locked chains with two pieces.
Back in January~1998, the third author showed how to simulate this behavior
with convex shapes, indeed, just three triangles;
see Figure \ref{locked 3 triangles}.
This example has circulated throughout the years to many researchers
(including the authors of this paper) who have asked about chains of 2D shapes.
The only really unsatisfying feature of the example is that some of the angles
are very obtuse.  But with a little more work, one can find examples with
acute angles, indeed, equilateral triangles, albeit of different size;
see Figure \ref{locked unequal equilateral}.  What could be better than
equilateral triangles?

It is therefore reasonable to expect, as we did for many years, that
there is no interesting class of shapes, other than line segments,
with a universality result---essentially all other shapes admit locked chains.
We show in this paper, however, that this guess is wrong.

We introduce a family of shapes, called \emph{slender adornments},
and prove that all open 
chains, made up of
arbitrarily many different shapes from this family,
can be universally reconfigured between any two states.
Indeed, we show that these chains have a natural canonical configuration,
analogs of the straight configuration of an open chain. 
Our result is based on the existence of 
``expansive motions'', proved in
\cite{Connelly-Demaine-Rote-2003}.
Our techniques build on the theory of unfolding chains of line segments,
substantially generalizing and extending the results from that theory.  Indeed, 
the results in this paper essentially piggy-back onto the results of \cite{Connelly-Demaine-Rote-2003} or any of the other results and algorithms such as \cite{Streinu-2000,Streinu-2005}
 that provide a continuous expansive motion of the base chain.  
Our results go far beyond what \emph{we} thought was possible
(until recently).   As part of the methods that we describe here, we also consider discrete expansive motions of the base chain, that do not necessarily come from a continuous expansive motion.  (A \emph{discrete expansion} of a chain $C$  is simply another corresponding chain $C'$ such that if $x$ and $y$ are two points in $C$ the distance between corresponding points $x'$ and $y'$ is not smaller than the distance between $x$ and $y$.)   In that case if all the slender adornments are \emph{symmetric} under the reflection about the line of the base chain, then any expansive discrete motion of the base chain will have the property that the attached adornments will not overlap.  It turns out that the continuous case, when the adornments are not necessarily symmetric, follows from the discrete symmetric case.  

The family of slender adornments has several equivalent definitions.
The key idea is to distinguish the two hinge points on the boundary
of the shape connecting to the adjacent shapes in the chain,
and view the shape as an \emph{adornment} to the line segment connecting
those two hinges, called the \emph{base}.  This view is without loss of
generality, but provides additional information relating the shapes and
how they are attached to neighbors,
which turns out to be crucial to obtaining a universality result.
An adornment is \emph{slender} if the distance from either endpoint of the base
to a point moving along one side of the adornment
changes monotonically.  If the boundary curve, defining the adornment,
is sufficiently smooth, it is slender if and only if every inward
normal of the shape hits the base.  Equivalently, an adornment is
slender if it is the union, in each half-plane having the base as a
boundary, of the intersection of pairs of disks centered at the two
endpoints of the base.
For the first and last edge of an open chain, there may be some freedom in defining the base.

\begin{figure}[htbp]
  \centering
  \includegraphics[width=\linewidth]{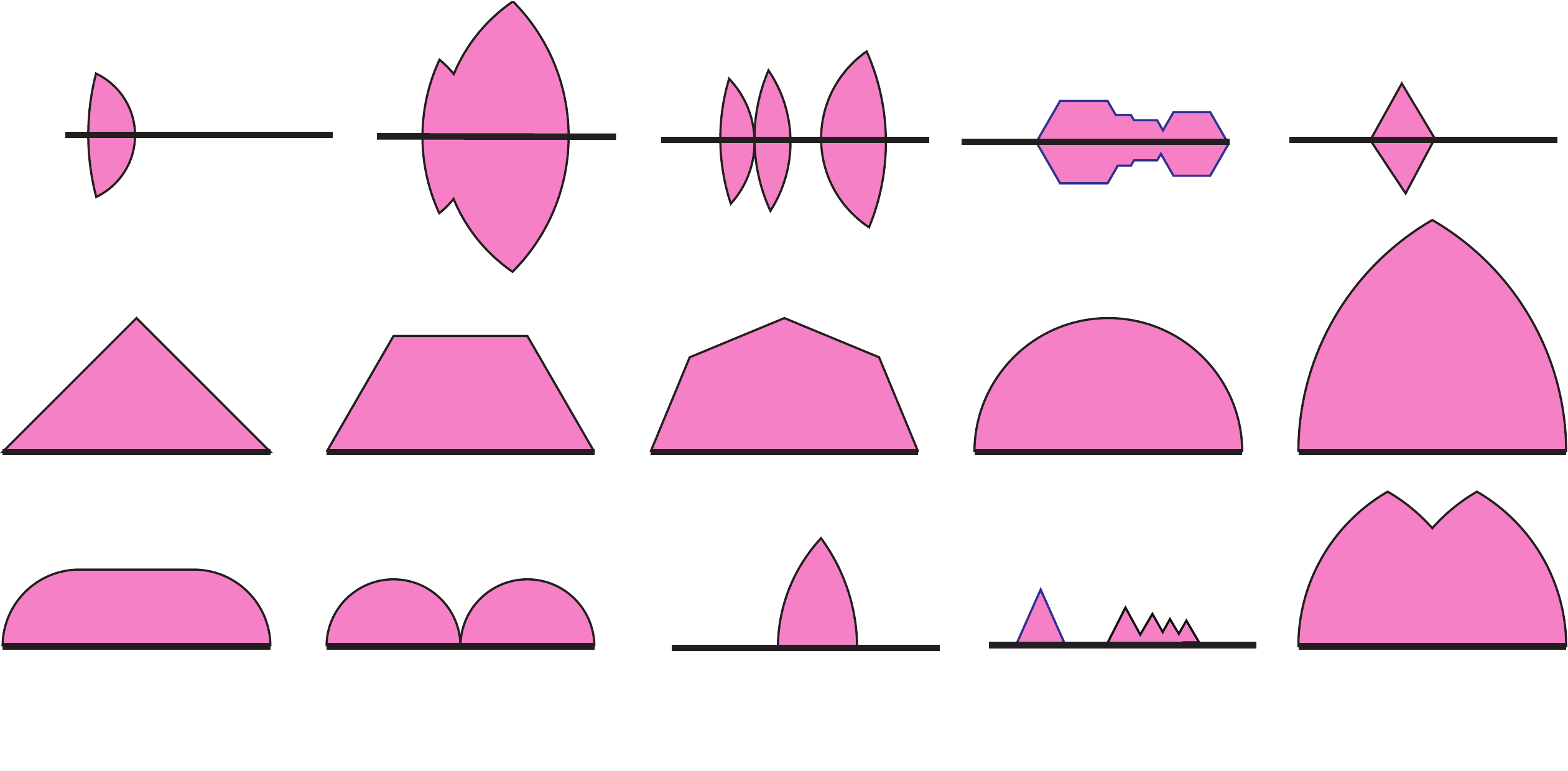}
  \caption{Examples of slender adornments.  The base is drawn bold. The examples in the top row are symmetric. Any two of these examples can be glued together along a common base so that the union also becomes a slender adornment.}
  \label{slender}
\end{figure}

Slender adornments are quite general.
Figure \ref{slender} shows several examples of slender adornments.
These examples are themselves slender adornments, but also any pair can be
joined along their bases so that the union makes another slender adornment.
Our results imply that one can take any of these slender adornments,
link the bases together into an open chain in any way that the chain does not
self-intersect, and the resulting chain can be unfolded without
self-intersection to a straight  configuration,
and thus the chain can be folded without self-intersection
into every configuration.

We also demonstrate the tightness of the family of slender adornments by
giving examples of locked chains of shapes that are not slender.
Specifically, we show that, for any desired angle $\theta < 90^\circ$,
there is a locked chain of isosceles triangles with apex angle~$\theta$.
This is precisely the family of isosceles-triangle adornments
that are not slender.
Thus, for chains of triangles, obtuseness is really desirable,
contrary to our intuition from Figure~\ref{locked 3 triangles}:
the key is that the apex angle opposite the base (in the adornment view)
be nonacute, not any other angle.
The proof that our examples are locked uses the self-touching theory
developed for trees of line segments
in \cite{Connelly-Demaine-Rote-2002-infinitesimally-locked}.

\begin{figure}[htbp]
  \centering
  \includegraphics[width=0.5\linewidth]{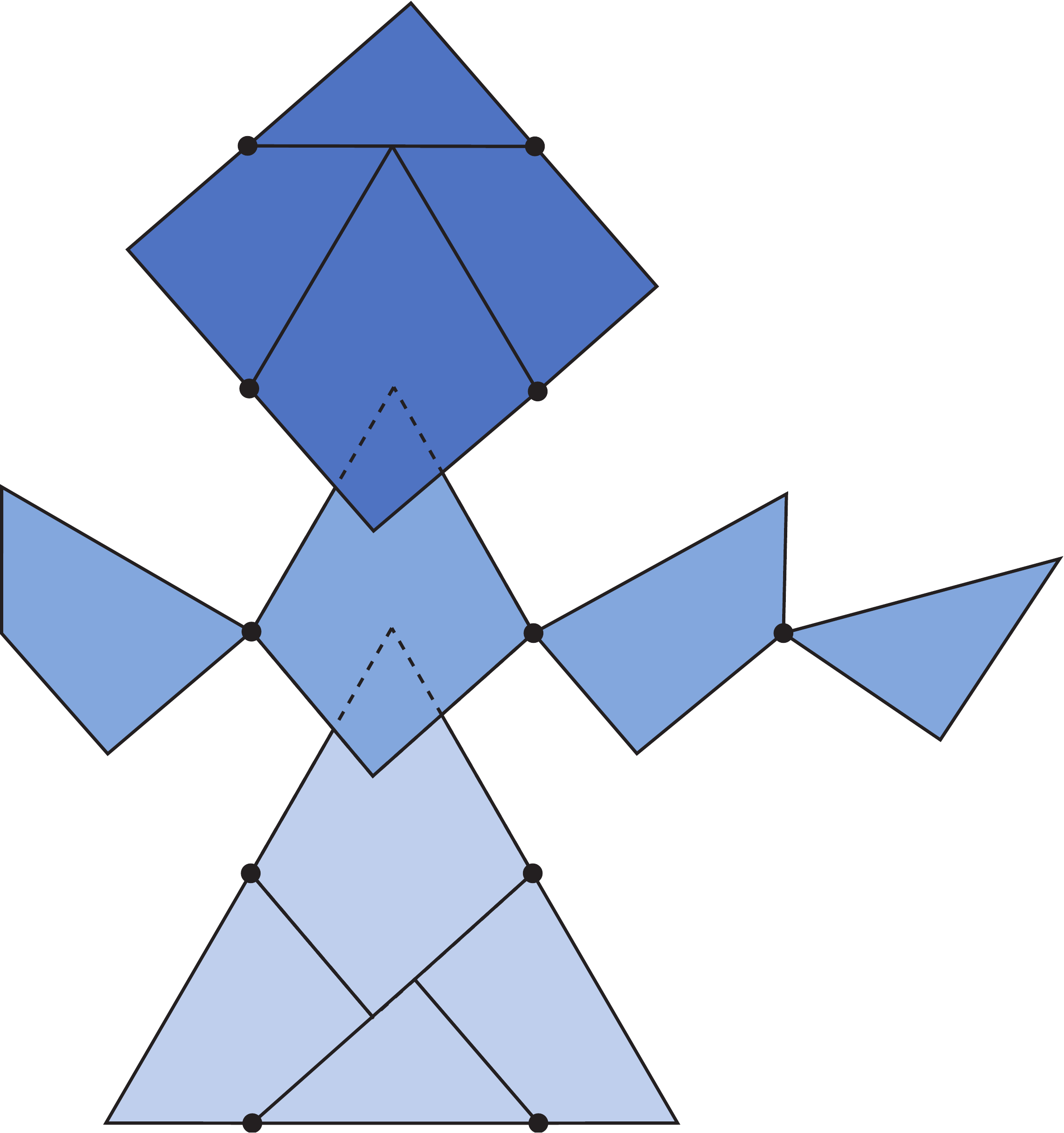}
  \caption{Hinged dissection of square to equilateral triangle,
    described by Dudeney~\protect\cite{Dudeney-1902-hinged}.
    Different shades show different folded states (overlapping slightly).}
  \label{dudeney}
\end{figure}

\paragraph{Motivation.}
Hinged collections of rigid objects have been studied
previously in many contexts, particularly robotics.
One recent body of algorithmic work by Cheong et al.~\cite{csgor-ihp-06}
considers how chains of polygonal objects can be \emph{immobilized}
or \emph{grasped} by a robot with a limited number of actuators.
Grasping is a natural first step toward robotic \emph{manipulation},
but the latter challenge requires a better understanding of reconfigurability.
This paper offers the first theoretical underpinnings for reconfiguration
of chains of rigid objects (other than line segments).

Another potential application is to continuous folding of
hinged dissections.  Hinged dissections are chains or trees of polygons
that can be reconfigured into two or more self-touching configurations
with desired silhouettes.  For example, Figure~\ref{dudeney} shows
a classic hinged dissection from 1902 of a square into an equilateral
triangle of the same area.  Many general families of hinged dissections
have been established in the recent literature
\cite{Akiyama-Nakamura-1998,
Demaine-Demaine-Eppstein-Frederickson-Friedman-2005,
Demaine-Demaine-Lindy-Souvaine-2005,
Eppstein-2001-mirror-dissection,
Frederickson-2002}.
One problem not addressed in this literature, however, is whether the
reconfigurations can actually be executed without self-intersection,
as in Figure~\ref{dudeney}.
Our results provide potential tools, previously lacking,
for addressing this problem.
While hinged dissections have frequently been considered in recreational
contexts, they have recently found applications in nanomanufacturing
\cite{Mao-Thallidi-Wolfe-Whitesides-Whitesides-2002} and reconfigurable
robotics \cite{Demaine-Demaine-Lindy-Souvaine-2005}.

\paragraph{Outline.}
This paper is organized as follows.
Section \ref{Slender Adornments} defines the model and slender adornments
more precisely, and proves several basic properties.
Section \ref{Symmetric Adornments} describes the case when each adornment is symmetric about its base and is important for proving,
in Section \ref{Slender Adornments Cannot Lock},  that simple chains of
slender adornments can always be unfolded so that the base is convex or straight. In Section \ref{generalizations} we discuss the situation when the adornments are permitted to overlap.
Section \ref{locked sharp} describes our examples of locked chains of
isosceles triangles, including the necessary background
from self-touching trees.  The 
conclusion (Section~\ref{sec:conclusion})  includes several closing remarks.

The results of this paper have been reviewed and sketched in the
survey~\cite[Section~4]{Bobs-survey}, which cites this paper (in its
full form) as the source of the results and the proofs.

\section{Slender Adornments}
\label{Slender Adornments}

This section provides a formal statement of the objects we consider---adorned
chains consisting of slender adornments---and proves several basic results
about them.

\subsection{Adorned Chains}

Our object of study is a chain of nonoverlapping rigid planar shapes
(Jordan regions) that are hinged together sequentially at rotatable joints.
Another way to view such a chain is to consider the \emph{underlying polygonal
chain}, the \emph{core}, of line segments connecting successive joints.
(For an open chain, there is some freedom in choosing the endpoints for the
first and the last bar of the chain.)
On the one hand, these line segments can be viewed as \emph{bars} that move
rigidly with the shapes to which they belong.
On the other hand, the shapes can be viewed as ``adornments'' to the bars
of an underlying polygonal chain.
This view leads to the concept of an ``adorned polygonal chain'',
which we now proceed to define more precisely.

An \emph{adornment} is a simply connected compact region in the plane,
called the \emph{shape},
together with a line segment $x y$ connecting two boundary points,
called the \emph{base}.  There are two \emph{boundary arcs} from $x$
to $y$ that enclose the shape, called \emph{sides}.
We require the base to be contained in the shape; i.e., the base must be a chord of the shape.

We say that two distinct adornments \emph{overlap} when some point of one adornment lies in the interior of the other, and we insist that the relative interiors of the base chains be disjoint.  
Thus, the bases of two shapes are not allowed to touch except at common hinges of the polygonal chain.  An \emph{adorned polygonal chain} is a set of nonoverlapping adornments whose
bases form a polygonal chain.     We permit the shapes to touch on their boundary and to slide along each other.

For our main result, Theorem \ref{thm:main}, where we assume that the motion of the base is expansive, it is not necessary to assume that the base chain is simple.   It can be any finite embedded graph with straight edges whose relative interiors are pairwise disjoint; a vertex may touch an edge.  When the base chain is simple the results of \cite{Connelly-Demaine-Rote-2003} or \cite{Streinu-2005} guarantee that there is such an expansive motion.  On the the other hand, although an expansive motion of the base chain of a strictly simple closed polygon to a convex convex configuration can be guaranteed, it may happen that two realizations are not in the same configuration component, as shown in Figure \ref{fig:2-components}, and in the conclusion (Section~\ref{sec:conclusion}) there is a description of a case when there are infinitely many components in the configuration space.  

The viewpoint of a chain of shapes as an adorned polygonal chain is useful
for two reasons.
First, we can more easily talk about the kinds of shapes,
and their relation to the locations of the incident hinges,
in a family of chains: this information is captured by the adornments.
Second, the underlying polygonal chain provides a mechanism for folding
the chain of shapes, as well as a natural \emph{unfolding} goal:
straighten the underlying open chain or convexify the underlying closed chain.
Indeed, we show that, in some cases, unfolding motions
of the polygonal chain induce valid unfolding motions of the chain of shapes.

\subsection{Slender Adornments}

\begin{figure} [here]
\begin{center}
\includegraphics{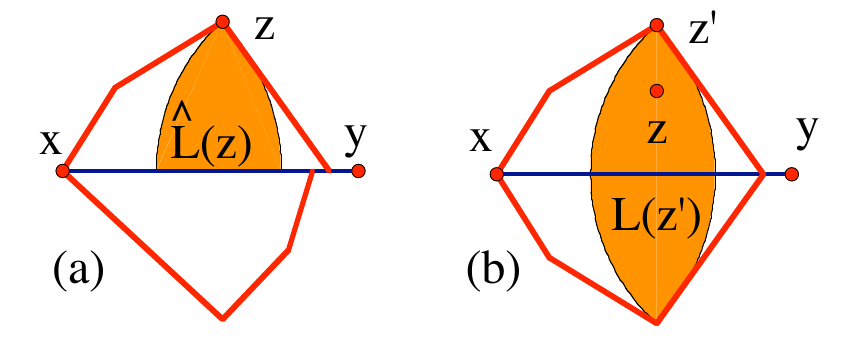}
\caption{ (a) a half-lens in non-symmetric adornment. (b) a symmetric lens with a point $z$ in the interior of the lens and the adornment.}
 \label{fig:lenses}
\end{center}
\end{figure}

An adornment is defined to be \emph{slender} if,  for a point moving on either side of
  the shape, the distance to each endpoint of the base changes monotonically (possibly not strictly
  monotonically).  An adornment is called \emph{symmetric} if it is symmetric about the line through the 
  base.  An adornment is called \emph{one-sided} if it lies in just one of the closed half-planes whose 
  boundary contains the base.  Clearly, a general adornment is the union of two one-sided adornments, and a one-sided adornment is the intersection of a symmetric adornment with a closed half-plane whose boundary contains the base.  For a base segment $[x,y]$ and a point $z$ in the plane, where $\|z-x\| \le \|y-x\|$ and  $\|z-y\| \le \|x-y\|$, let $L(z)$ be the intersection of the two disks with $z$ on its boundary, centered at $x$ and 
 at $y$.  We call $L(z)$ the \emph{lens determined by $z$ associated to the base $[x, y]$}.  A \emph{half-lens}, denoted as $\hat{L}(z)$, is the intersection of $L(z)$ and the closed half-plane through the base containing $z$.   See Figure \ref{fig:lenses} for a picture of a half-lens and lens. 
The following are some simple, but useful, properties of lenses.  

\begin{figure}[htbp]
  \centering
  \begin{overpic}[scale=0.5]{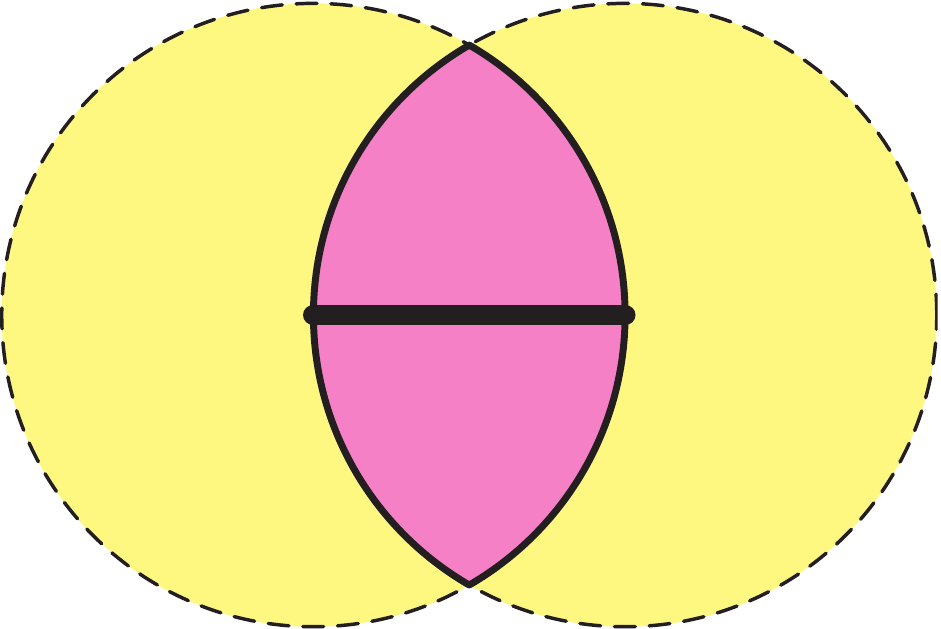}
    \put(48,65){$z$}
    \put(26,31.5){$x$}
    \put(70,32){$y$}
  \end{overpic}
  \caption{The largest slender adornment with a given base is a lens $L(z)$, where $\|z-x\| =  \|z-y\| = \|y-x\|.$}
  \label{crescent}
\end{figure}

\begin{proposition} \label{prop:lenses} For any point $z$ in a \textup(symmetric\textup) slender adornment $A$,  $\hat{L}(z) \subset A$ \textup($L(z) \subset A$\textup). 
\end{proposition}
\begin{proof}  Let $z$ be a point on the defining boundary of  $A$. Since the distance to $x$ along the boundary is monotone, no point along the path from $z$ to $y$ intersects the interior of the circle centered at $x$ through $z$.  Similarly, no point along the path from $z$ to $x$ intersects the interior of the circle centered at $y$ through $z$.  Thus, the intersection of the circular disks  centered at $x$ and $y$, with $z$ on their boundary, and the closed half-plane containing $z$, $\hat{L}(z)$ is contained in the adornment.  In the symmetric case, the intersection of the circular disks with $z$ on their boundary $L(z)$ is contained in $A$. See Figure \ref{fig:lenses}.
\end{proof}
\begin{proposition} \label{prop:lens-interior} For any point $z$ in the interior of a \textup(symmetric\textup) slender adornment $A$, there is a half-lens $\hat{L}(z') \subset A$ \textup(lens $L(z') \subset A$\textup) that has $z$ in its interior.
\end{proposition}
\begin{proof}  The half-lens (lens) through $z$ is contained in $A$ by Proposition \ref{prop:lenses}.  Since $z$ is in the interior of $A$, there is another point $z'$ in $A$ on the line perpendicular to the base segment slightly further away from the base.  Then $z$ is in the interior of the half-lens (lens) defined by $z'$.  
\end{proof}
\begin{proposition} \label{prop:lens-union} A symmetric adornment $A$
  of a base $[x,y]$ is slender if and only if it is the union of the
  intersection of pairs of disks centered at $x$ and $y$.
 \end{proposition} 
 \begin{proof}
Assume that $A$ is a slender adornment.
 By Proposition \ref{prop:lenses}, 
 the union of the lenses $L(z)$ for $z$ on the boundary of $A$ is
 contained in~$A$.
  
 To show the reverse containment, any point $z$ in the interior of $A$
 lies on a circle centered at $x$, and this circle must intersect the
 boundary of $A$ in (at least) one point $z'$.  Then $z$ is
 in $L(z')$.  Thus, the union of the lenses $L(z')$ for $z'$ on the
 boundary of the slender adornment contains $A$.
 
For the converse implication,
assume that we have a symmetric adornment $A$ which is formed as the union of lenses,
as stated in the proposition.
 We first show that for two points $z$ and $z'$ on
 the boundary of $A$,
we cannot have
  $\|z-x\| < \|z'-x\|$ and
  $\|z-y\| < \|z'-y\|$.
If this were the case, some lens on whose boundary $z'$ lies would contain $z$ in its interior, a contradiction.
Since we can exchange the role of $z$ and $z'$, we conclude
\begin{equation}
  \label{two-boundary-points}
  \|z-x\| < \|z'-x\| \implies   \|z-y\| \ge \|z'-y\|
\text{, \ and \ }
  \|z-x\| > \|z'-x\| \implies   \|z-y\| \le \|z'-y\|
\end{equation}
Now, if we order the points $z$ on one boundary chain from $x$ to $y$ by the quantity
$\|z-x\|-\|z-y\|$, we conclude that the distance
$\|z-x\|$ cannot decrease and the distance $\|z-y\|$ cannot increase; otherwise we would derive a contradiction to~\eqref{two-boundary-points}.
 \end{proof}
\begin{proposition} Finite unions and arbitrary intersections of slender adornments are slender adornments.
   \end{proposition} 
   \begin{proof} This follows from Proposition \ref{prop:lens-union} in the symmetric case, and the non-symmetric case follows from the symmetric case by intersecting with the closed half-plane containing the line segment.
   \end{proof}
 \begin{proposition} Every slender adornment is contained in the symmetric lens determined by either of the points equidistant from the endpoints of the base as in Figure \ref{crescent}.
    \end{proposition} 
   \begin{proof} Any slender adornment must be contained in the disk through the other end of the base, and thus it is in the intersection of those two disks.
   \end{proof} 

\subsection{Alternate Definitions of Slender Adornments}

\begin{figure}[h!]
  \centering
  \vspace*{-2ex}
  \includegraphics{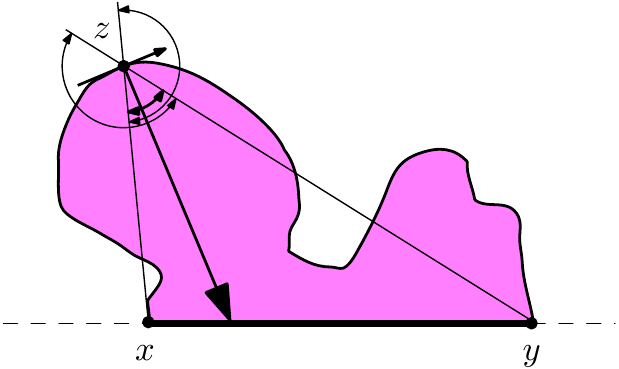}
  \caption{The normal property for slender shapes.}
  \label{slender versus monotone}
\end{figure}

In the preliminary conference version of this paper~\cite{conf-version}, we used
a more restricted definition of slenderness:
suppose that the boundary of an adornment consists of two
differentiable curves between the base points.
Then the condition of being slender is equivalent to requiring that every inward normal of the shape intersects the base before exiting the shape, as 
 illustrated in Figure~\ref{slender versus monotone}.
This
property ensures that slender adornments will not get closer together during an
expansive motion of the base.  Our current slenderness definition by
the monotone distance property is more general, easier to handle, and
it does not raise questions of differentiability.

\subsection{Kirszbraun's Theorem}
\label{Kirszbraun}

In what follows it is very handy to have the following theorem of Kirszbraun \cite{Kirszbraun-1934}.  

\begin{theorem}\label{thm:Kirszbraun} Suppose a finite set of closed circular disks in Euclidean space are rearranged so that no pair of centers gets strictly closer together.  If the original set has an empty intersection, so does the rearranged set.
\end{theorem}

There is a discussion and proof of this in \cite{Alexander-1984} as well as references to other proofs.  We only need this result for four disks in the Euclidean plane.

\section{Expanded Slender Symmetric Adornments Never Overlap}
\label{Symmetric Adornments}

We first prove the following for the case of symmetric slender adornments.  Note that the following result is for discrete expansions of the base chain.  Recall that two adornments \emph{overlap} if a point in one adornment lies in the interior of the other.  This allows their boundaries to touch, but not to penetrate each other.  Note that the bases of a chain do not cross as well, by the expansive property of a discrete motion.   We do not need the continuous expansive property for this result.
 
\begin{figure} [here]
\begin{center}
\includegraphics{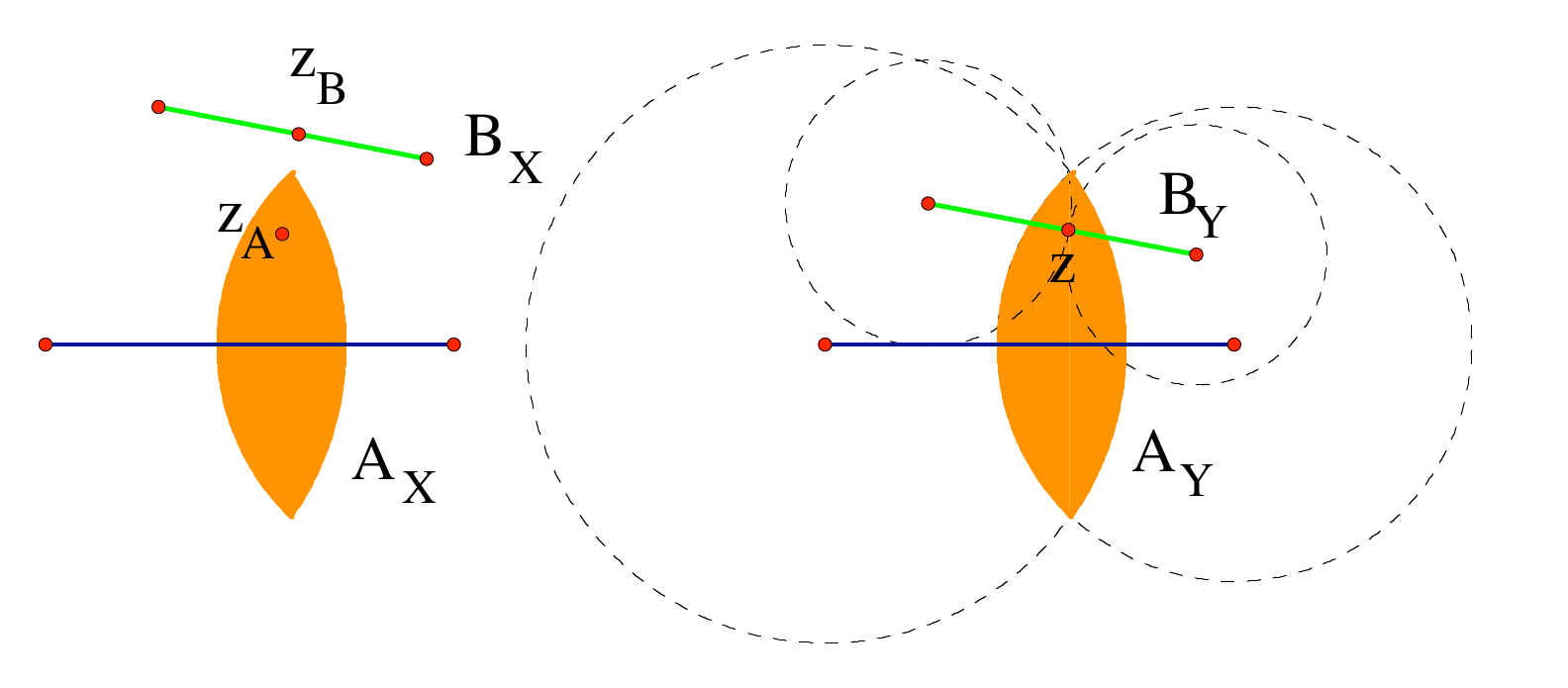}
\caption{This is the situation when two adornments overlap.  The four circles that used in the application of Kirszbraun's Theorem are indicated.  Note that, in this figure,  the motion from $X$ to $Y$ is not an expansion, since that would contradict Theorem \ref{thm:symmetric}.} \label{fig:symmetric-adornments}
\end{center}
\end{figure}

\begin{theorem}\label{thm:symmetric} Consider two configurations $X$ and $Y$ of corresponding chains  with symmetric slender adornments such that the base chain of  $Y$ is an expansion of the base chain of $X$.  We assume that the adornments attached to the base chain of $X$ do not overlap.  Then, when the corresponding adornments are attached to the base chain of $Y$, they also do not overlap.  
\end{theorem}
\begin{proof}  Suppose $A_X$ and $B_X$ are two slender adornments attached to different links of the base chain of $X$, and $A_X$ and $B_X$ do not overlap.   Let $A_Y$ and $B_Y$ be the  corresponding adornments for $Y$.  Suppose that $z$ is a point in the intersection $A_Y \cap B_Y$, where $z$ is in the interior of, say, $A_Y$.  We wish to find a contradiction.  

Let $z_A$ and $z_B$ be the corresponding distinct points in $A_X$ and $B_X$, respectively, that map to $z$ under the expanding map of their bases.  Thus, the lenses $L_A(z_A)$ and $L_B(z_B)$ for $A_X$ and $B_X$ have disjoint interiors, since the adornments do not overlap. Since $z$ is in the interior of  $A_Y$, we can assume that $L_A(z_A)$ can be chosen so that the closed lenses $L_A(z_A)$ and $L_B(z_B)$ are disjoint also. Thus the four circular disks that correspond to the circular disks that define  $L_A(z_A)$ and $L_B(z_B)$ have an empty intersection.  By Kirszbraun's Theorem \ref{thm:Kirszbraun}, and the expansion property of the endpoints of the bases of $A_X$ and $B_X$, which are the centers of the four circular disks, the intersection of the corresponding lenses for  $A_Y$ and $B_Y$ must also be empty, contradicting the assumption that  $A_Y$ and $B_Y$ overlap.  See Figure \ref{fig:symmetric-adornments}.
\end{proof}

\begin{figure} [here]
\begin{center}
\includegraphics{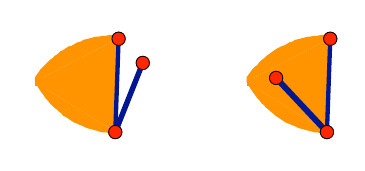}
\caption{This shows two chains, with slender but not symmetric adornments, where one is an expansion of the other, while there is an overlap in the expanded configuration, but not the original.}
 \label{fig:non-symmetric-overlap}
\end{center}
\end{figure}

For discrete expansions, it is not possible to deal with non-symmetric adornments.  Figure \ref{fig:non-symmetric-overlap} shows an example of two chains with corresponding slender adornments, one an expansion of the other.   One starts with no overlap, and the other has such an overlap.

\section{Slender Adornments Cannot Lock}
\label{Slender Adornments Cannot Lock}

We now consider the general case, assuming a continuous expansive motion. 

\begin{theorem} \label{thm:main} Suppose there is a continuous expansive motion of the base chain with slender non-overlapping, not necessarily symmetric, adornments attached.   Then the adornments never overlap during the motion.
\end{theorem}
\begin{proof}Because of the expansive property, two segments of the base chain can only intersect at common endpoints of adjacent segments.   Thus, suppose $z_A$, in the interior of adornment $A$, intersects $z_B$ in adornment $B$ at some time $t_1$ during the motion.  We look for a contradiction.  By Proposition \ref{prop:lens-interior}, there is a closed half-lens $L_A$ for $A$ that contains $z_A$ in its interior and there is a first time $t_0 < t_1$ when $L_A$ intersects another half-lens $L_B$ for $B$ that contains $z_B$.  Necessarily, that intersection must be on the common boundary of $L_A$ and $L_B$.  (Note that $L_B$ could be a single point on a base segment.)  Then there are three cases that can occur.  In each case, we will show that when the motion is continued from $t_0$ to $t_1$, $z_A$ and $z_B$ cannot intersect.
\begin{enumerate} [{Case} 1:]
\item The bases of $A$ and $B$ intersect in the interior of at least one of the bases.  This cannot happen because the bases are initially disjoint and the motion is expansive.  See Figure \ref{fig:overlap}(a).
\item The base of $A$ or $B$ intersects the half-lens of the other.  The half lens can be extended to a full symmetric lens without overlaping the base of the other.  Applying Theorem \ref{thm:symmetric} we see that $z_A$ and $z_B$ cannot intersect upon further expansion.   See Figure \ref{fig:overlap}(b).
\item The half lenses of $A$ and $B$ intersect.  In this case both half lenses can be extended to non-overlapping symmetric lenses.  Again we apply  Theorem \ref{thm:symmetric} to see that $z_A$ and $z_B$ cannot intersect upon further expansion.  See Figure \ref{fig:overlap}(c).
\end{enumerate}
 \begin{figure} [here]
\begin{center}
\includegraphics{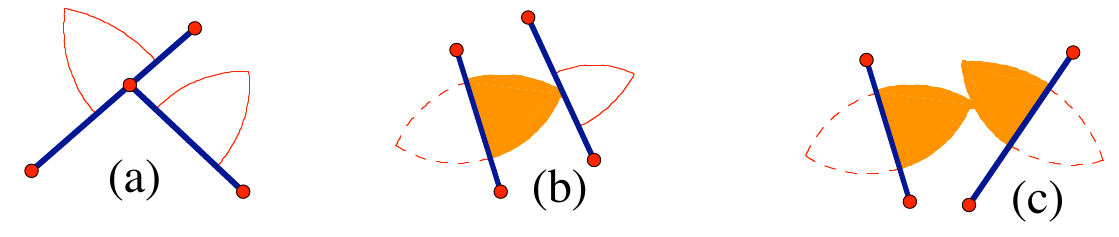}
\caption{This presents the cases when one adornment with its base might start to overlap with the other.  The dashed lines indicate where one or both of the lens of the adornment can be extended so that it does not intersect the relevant part of the other.  The thick lines indicate the part of the adornment that is not to be penetrated by the other lens or base. The thin lines indicate where some of the rest of the adornment might lie, containing the point $z_A$, say, in the proof.} \label{fig:overlap}
\end{center}
\end{figure}
\end{proof}

\begin{corollary}\label{cor:simple-chains}  A strictly simple polygonal chain with slender adornments attached can always be straightened or convexified by a continuous motion.  
\end{corollary}
\begin{proof} By  \cite{Connelly-Demaine-Rote-2003}, there is a continuous expansive motion of the base chain, where the final configuration is convex in the case of a closed chain and straight in the case of an open chain.  Then Theorem \ref{thm:main} implies that they can be carried along without overlap.
\end{proof}

\begin{corollary} \label{cor:open-chains} A strictly simple open
  polygonal chain with slender adornments
  can be continuously reconfigured between any
  two states.
\end{corollary}
\begin{proof} By Corollary \ref{cor:simple-chains}, both chains can be continuously expansively reconfigured so that the base chains are straight.
 Thus, one state can be expanded to have a straight base configuration, and then contracted to the other configuration by running its expansion backwards.
\end{proof}

\begin{figure} [here]
\begin{center}
\includegraphics{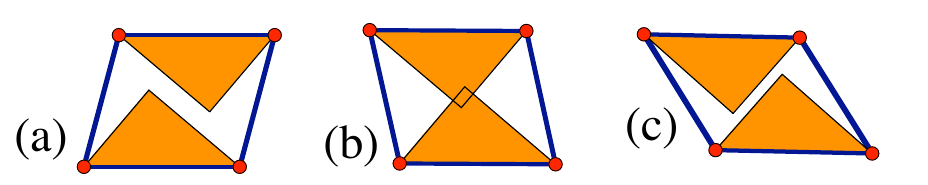}
\caption{Two configurations (a) and (c) of a quadrilateral with two
  slender adornments attached. It is not possible to continuously move
  from one to the other without colliding. Figure (b) shows how the
  two adornments collide as the quadrilateral is deformed from (a) to
  (c). }
 \label{fig:2-components}
\end{center}
\end{figure}

It is interesting to note that the conclusion of Corollary \ref{cor:open-chains} does not hold for closed chains, even though any two convex chains with no adornments  can be continuously reconfigured from one to the other.  Figure \ref{fig:2-components} shows an example, where the configuration space has two components, where the base chain is a quadrilateral,  and where each adornment is a triangle attached to its base. 

Indeed, in the conclusion (Section~\ref{sec:conclusion}) it is shown how to create a quadrilateral with two slender adornments such that the configuration space has infinitely many components.  

\begin{figure} [here]
\begin{center}
\includegraphics{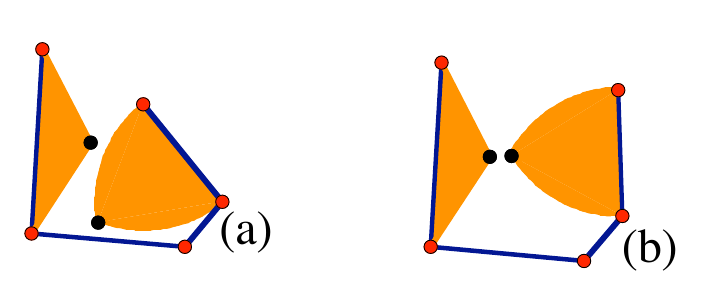}
\caption{The base chain of Figure (a) expands to Figure (b).  But the dark points on the corresponding slender adornments get closer together.}
 \label{fig:non-expansive}
\end{center}
\end{figure}

It is also interesting to note that
when the base chain is expanded, it often happens that the motion on the
adorned configuration is not expansive.
Figure~\ref{fig:non-expansive} shows an example.

\section{Generalizations:  Overlapping Adornments and Generalized Slender Symmetric Adornments}\label{generalizations}

In the discussion so far, we have assumed, when the adornments are attached to their chains, that they do not overlap. What happens when the slender adornments do overlap?  It turns out that we can apply some of the results of \cite{Bezdek-Connelly-2002} related to problems concerning areas of unions and intersections of circular disks in the plane to the case when the adornments are all symmetric. 

Proposition \ref{prop:lens-union} shows that any symmetric adornment is the infinite union of symmetric lenses $L(z)$ for all $z$ on the boundary of the adornment.  To apply the theory of \cite{Bezdek-Connelly-2002} it is more convenient that there only be a finite number of sets involved in the union of lenses.  But it is easy to see that each adornment can be approximated by a finite union of lenses.  

We first define a \emph{flower} as a set in the plane that can be described in terms of finite unions and intersections of circular disks, where each disk appears once and only once in the Boolean expression that describes the set.  For the special case at hand we need only be concerned with flowers $F$ of the following sort:
\begin{equation}
F=(B_1 \cap B_2) \cup (B_3 \cap B_4) \cup (B_5 \cap B_6) \cup \dots \cup (B_{N-1} \cap B_{N}) ,\label{flower}
\end{equation}
where each $B_i, i=1, \dots, N$ is a circular disk in the plane.  Flowers were defined by \cite{Gordon-Meyer-1995}, and a special case of Corollary 8 in \cite{Bezdek-Connelly-2002} shows the following.  Let $B(x, r)$ denote a disk in the plane of radius $r$ centered at $x$.
\begin{lemma}\label{monotone-area}
Let $B(p_i, r_i)$ and $B(q_i, r_i), i= 1, \dots, N$ be two sets of planar disks, where $\|p_i-p_{i+1}\| \ge \|q_i-q_{i+1}\|$ for $i$ odd, and $\|p_i-p_{i+1}\| \le \|q_i-q_{j}\|$ for all other pairs $i<j$.  Then the area of the flower $F$ in \eqref{flower} defined for the configuration of  $p_i$ is less than or equal to the area of the flower $F$ defined for the configuration of  $q_i$.  
\end{lemma}  

The crucial observation is that the union of the slender adornments can be approximated by flowers.  Each lens is the intersection of two disks, one of the terms in  (\ref{flower}), and Proposition \ref{prop:lens-union} implies the following.

\begin{figure} [here]
\begin{center}
\includegraphics{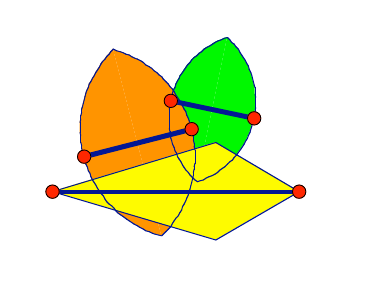}
\caption{Three intervals with overlapping symmetric slender adornments. }
 \label{fig:area-overlap}
\end{center}
\end{figure}

\begin{figure} [here]
\begin{center}
\includegraphics{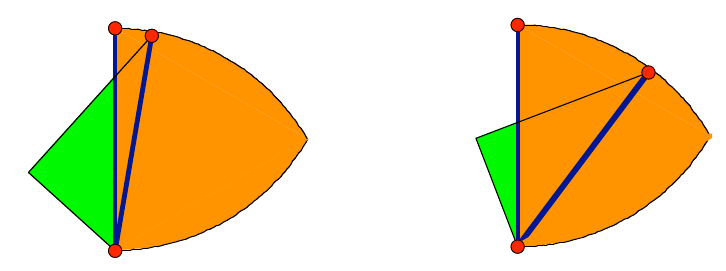}
\caption{An example of two chains with non-symmetric slender adornments, where the expanded chain with adornments has smaller area.}
 \label{fig:area-down}
\end{center}
\end{figure}

\begin{theorem} \label{thm:overlap} Suppose that one chain is a discrete expansion of the other and slender symmetric adornments are attached to each chain.  Then the area of the union of the adornments does not decrease.  
\end{theorem}
Figure \ref{fig:area-overlap} shows an example of overlapping
symmetric adornments.  Figure \ref{fig:area-down} shows an example of
a chain with non-symmetric adornments that expands to another chain
and the area of the slender adornments with an expanded core
decreases.

Another possible generalization is to attach the analog of slender
adornments to simplicial complexes in higher dimensions.  For example,
a set $A$ in three-space would be called \emph{slender with respect to
  a triangle base $B$} if for any plane $P$ perpendicular to the plane
of $A$, $P \cap A$ is slender with respect to $B\cap A$.  Then the
analog of 
Theorem~\ref{thm:main}
 should hold using
the notion of symmetric slender adornments.  Even the analog of
Theorem~\ref{thm:overlap} for the volumes of symmetric slender
adornments would still hold, but it could only be asserted for
continuous expansions of the base chain.  The higher dimensional
version of Corollary 8 in \cite{Bezdek-Connelly-2002}, on which
Lemma~\ref{monotone-area} is based, is not known for discrete
expansions.  However, in \cite{Csikos-2001}, there is a continuous
version that will suffice.  In higher dimensions, the idea is to
assume simply that the base chain, to which the adornments are
attached, is expanded.

\section{Locked Chains of Sharp Triangles}
\label{locked sharp}

An isosceles triangle with an apex angle of $\geq 90^\circ$
and with the nonequal side as the base is a slender adornment.
By Corollary~\ref{cor:simple-chains}, any chain of such triangles
can be straightened.  In this section we show that this result is tight:
for any isosceles triangle with an apex angle of $< 90^\circ$
and with the nonequal side as a base, there is a chain of these triangles
that cannot be straightened.

\begin{figure}
  \centering
  \subfigure[]
    {\includegraphics[scale=0.6]{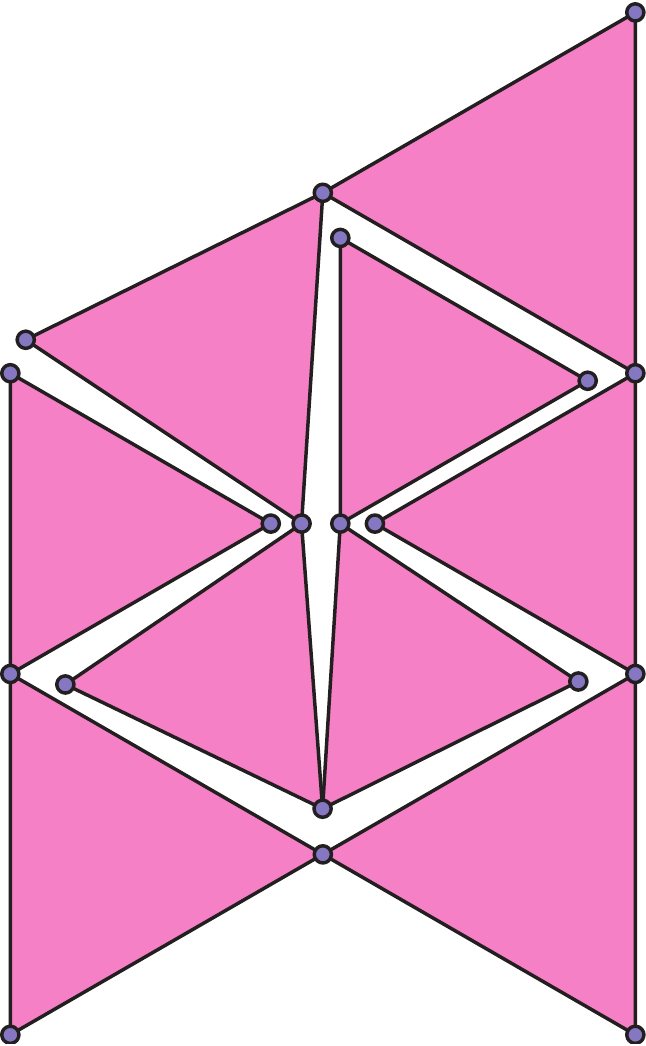}
     \label{locked 9 equilateral loose}}
  \hfil\hfil
  \subfigure[]
    {\includegraphics[scale=0.6]{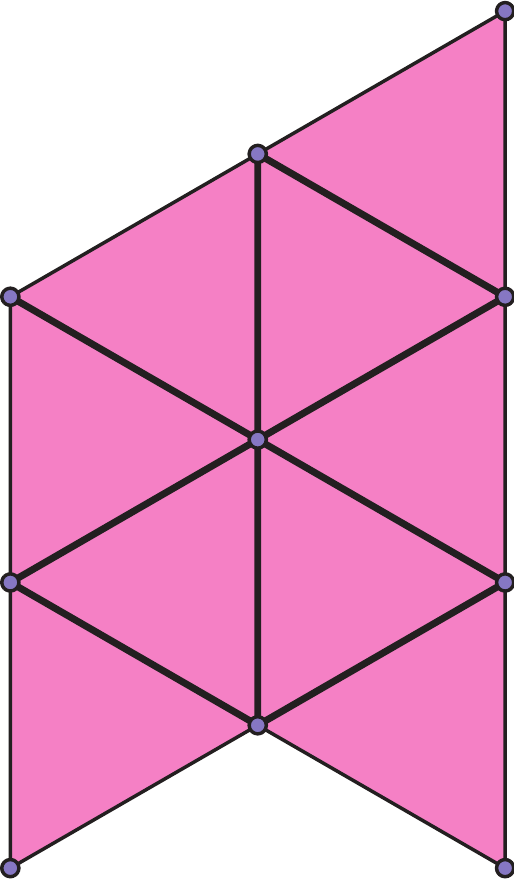}
     \label{locked 9 equilateral tight}}
  \caption{A locked chain of nine equilateral triangles.
    (a) Drawn loosely.  Separations should be smaller than they appear.
    (b) Drawn tightly, with no separation, as a self-touching configuration.}
  \label{locked 9 equilateral}
\end{figure}

\begin{figure}
  \centering
  \includegraphics[scale=0.6]{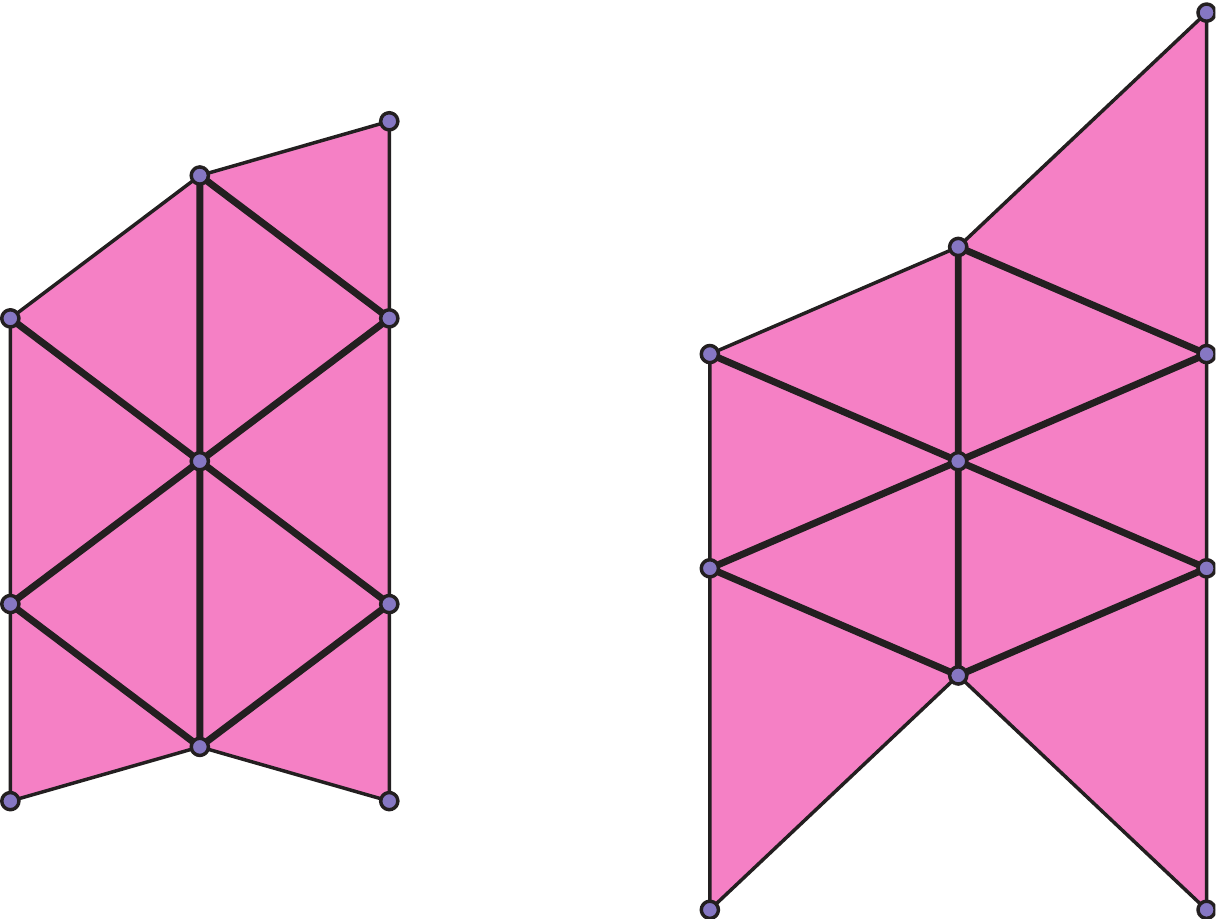}
  \caption{Variations on the self-touching configuration from
    Figure~\protect\ref{locked 9 equilateral tight}
    to have any desired angle $< 90^\circ$ opposite the base of each triangle.}
  \label{locked 9 isosceles tight}
\end{figure}

Figure~\ref{locked 9 equilateral loose} shows the construction for equilateral
triangles (of slightly different sizes).
This figure is drawn with the pieces loosely separated, but the
actual construction has arbitrarily small separations and arbitrarily closely
approximates the self-touching geometry shown in
Figure~\ref{locked 9 equilateral tight}.
Stretching the triangles in this self-touching geometry,
as shown in Figure~\ref{locked 9 isosceles tight},
defines our construction for any isosceles triangles with an opposite angle of
any value less than $90^\circ$.
In this case, however, our construction uses two different scalings
of the same triangle.

\subsection{Theory of Self-Touching Configurations}

This view of the construction as a slightly separated version of a
self-touching configuration allows us to apply the program developed in
\cite{Connelly-Demaine-Rote-2002-infinitesimally-locked} for proving a
configuration locked.  This theory allows us to study the rigidity of
self-touching configurations, which is easier because vertices
cannot move even slightly,
and obtain a strong form of lockedness of non-self-touching perturbations
drawn with sufficiently small (but positive) separations.

To state this relation precisely, we need some terminology from
\cite{Connelly-Demaine-Rote-2002-infinitesimally-locked}.
Call a linkage configuration \emph{rigid} if it cannot move at all.
Define a \emph{$\delta$-perturbation} of a linkage configuration to be
a repositioning of each vertex within distance $\delta$ of its original
position, without regard to preserving edge lengths
(better than $\pm 2 \delta$),
but consistent with the combinatorial information of which vertices are
on which side of which bar.
Call a linkage \emph{locked within~$\epsilon$} if no motion that leaves some
bar pinned to the plane moves any point by more than $\epsilon$.
Call a self-touching linkage configuration \emph{strongly locked} if,
for any desired $\epsilon > 0$, there is a $\delta > 0$ such that
all $\delta$-perturbations are locked within~$\epsilon$.
Thus, if a self-touching configuration is strongly locked, then
the smaller we draw the separations in a non-self-touching perturbation,
the less the configuration can move.  In particular, if we choose $\epsilon$
small enough, the linkage must be locked in the standard sense of having
a disconnected configuration space locally.

\begin{theorem}
  {\rm \latexcite[Theorem~8.1]{Connelly-Demaine-Rote-2002-infinitesimally-locked}}
  If a self-touching linkage configuration is rigid,
  then it is strongly locked.
  \label{strong lock}
\end{theorem}

Therefore, if we can prove that the self-touching configuration in
Figure~\ref{locked 9 equilateral tight} (and its variations in
Figure~\ref{locked 9 isosceles tight}) are rigid, then sufficiently small
perturbations along the lines shown in Figure~\ref{locked 9 equilateral loose}
are rigid.

The theory of \cite{Connelly-Demaine-Rote-2002-infinitesimally-locked} also
provides tools for proving rigidity of a self-touching configuration.
Specifically, we can study \emph{infinitesimal motions},
which just define the beginning of a motion to the first order.
Call a configuration \emph{infinitesimally rigid}
if it has no infinitesimal motions.

\begin{lemma}
  {\rm \latexcite[Lemma~6.1]{Connelly-Demaine-Rote-2002-infinitesimally-locked}}
  If a self-touching linkage configuration is infinitesimally rigid,
  then it is rigid.
  \label{rigid rigid}
\end{lemma}

\begin{figure}[htbp]
  \centering
  \begin{overpic}{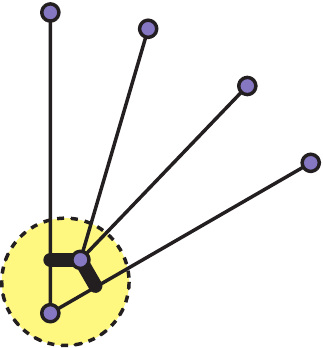}
    \put(10,10){\makebox(0,0)[br]{$v$}}
    \put(10,96){\makebox(0,0)[r]{$w_1$}}
    \put(90,46){\makebox(0,0)[t]{$w_2$}}
    \put(23,30){\makebox(0,0)[br]{$u$}}
  \end{overpic}
  \caption{Two zero-length connections between vertices $u$ and~$v$.}
  \label{zero length connection}
\end{figure}
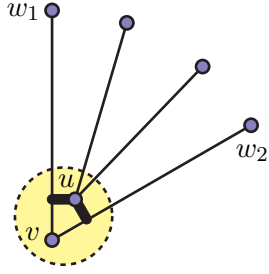

A final tool we need from
\cite{Connelly-Demaine-Rote-2002-infinitesimally-locked}
is for proving infinitesimal rigidity.
For each vertex $u$ wedged into a convex angle
between two bars $\{v,w_1\}$ and~$\{v,w_2\}$,
we say that there are two so-called \emph{zero-length connections} between $u$ and~$v$,
one perpendicular to each of the two bars $\{v,w_i\}$.
Such a connection between a point $u$ and a bar $\{v,w\}$ restricts $u$ to one side of the line through $v,w$,
\footnote{The definition of zero-length connections in
  \cite{Connelly-Demaine-Rote-2002-infinitesimally-locked} is more
  general, but this definition suffices for our purposes.  The term
  ``zero-length connection'' is perhaps unfortunate since it
  mistakenly suggests a constraint on the \emph{distance} between $u$
  and $v$ to be zero. The zero-length connection is, however, allowed
  to slide freely on the bar $\{v,w\}$.}
see Figure~\ref{zero length connection}.
These connections must increase to the first order because $u$ must not cross
the two bars $\{v,w_i\}$.
In proving infinitesimal rigidity, we can choose to discard any zero-length
connections we wish, because ignoring some of the noncrossing constraints
only makes the configuration more flexible.
Together, the bars and the zero-length connections are the \emph{edges}
of the configuration.
Define a \emph{stress} to be an assignment of real numbers (\emph{stresses})
to edges such that, for each vertex~$v$, the vectors with directions defined by
the edges incident to $v$, and with magnitudes equal to the corresponding
stresses, sum to the zero vector.
We denote the stress on a bar $\{v,w\}$ by~$\omega_{v w}$,
and we denote the stress on a zero-length connection between vertex $u$
and vertex $v$ perpendicular to $\{v,w\}$ by~$\omega_{u,v w}$.

\begin{lemma}
  {\rm \latexcite[Lemma~7.2]{Connelly-Demaine-Rote-2002-infinitesimally-locked}}
  If a self-touching configuration has a stress that is negative on every
  zero-length connection, and if the configuration is infinitesimally rigid
  when every zero-length connection is treated as a bar pinning two vertices
  together, then the self-touching configuration is infinitesimally rigid.
  \label{stress lemma}
\end{lemma}

\subsection{Locked Chains}

We are now in the position to state the precise senses in which the chains
of isosceles triangles in Figures \ref{locked 9 equilateral} and
\ref{locked 9 isosceles tight} are locked:

\begin{theorem} \label{triangles rigid}
  The self-touching chains of nine isosceles triangles shown in
  Figures \ref{locked 9 equilateral tight} and \ref{locked 9 isosceles tight}
  are rigid provided that the apex angle is $< 90^\circ$.
\end{theorem}

Applying Theorem \ref{strong lock}, we obtain the desired result:

\begin{corollary} \label{triangles locked}
  The self-touching chains of nine isosceles triangles shown in
  Figures \ref{locked 9 equilateral tight} and \ref{locked 9 isosceles tight}
  are strongly locked provided that the apex angle is $< 90^\circ$.
  Therefore, any sufficiently small non-self-touching perturbation,
  similar to the one shown in Figure \ref{locked 9 equilateral loose},
  is locked.
\end{corollary}

Sections \ref{Simplifying Rules}--\ref{Stress Argument}
prove Theorem \ref{triangles rigid}.

\subsection{Simplifying Rules}
\label{Simplifying Rules}

We introduce two obvious rules that significantly restrict the allowable motions
of the self-touching configuration of isosceles triangles.

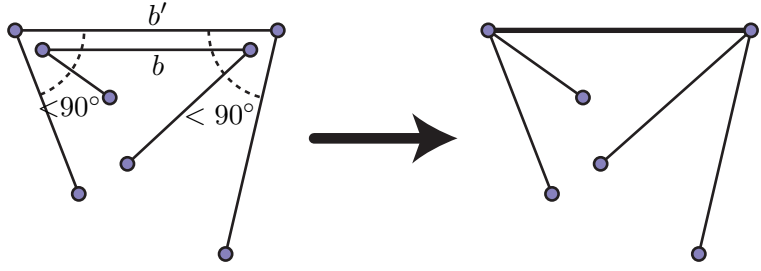
\begin{figure}
  \centering
  \begin{overpic}[scale=1]{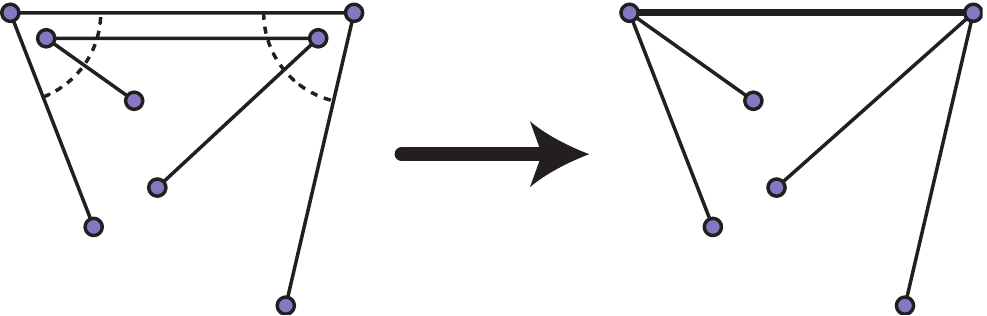}
    \put(20,31.5){\makebox(0,0)[b]{$b'$}}
    \put(20,27.5){\makebox(0,0)[t]{$b$}}
    \put(4,22){\makebox(0,0)[lt]{${<}90^\circ$}}
    \put(33,21){\makebox(0,0)[rt]{$< 90^\circ$}}
  \end{overpic}
  \caption{Rule~\protect\ref{rule 1}
           for simplifying self-touching configurations.}
  \label{rule1}
\end{figure}

\begin{onerule} \label{rule 1}
  If a bar $b$ is collocated with another bar $b'$ of equal length,
  and the bars incident to $b'$ form angles less than $90^\circ$
  on the same side as~$b$, then any motion must keep
  $b$ collocated with $b'$ for some positive time.
  See Figure~\ref{rule1}.
\end{onerule}

\begin{proof}
  The noncrossing constraints at the endpoints of $b$ and $b'$ prevent $b$
  from moving relative to $b'$ until the angles at the endpoints of $b'$
  open to $\geq 90^\circ$,
  which can only happen after a positive amount of time.

  More formally, suppose without generality of loss that the endpoints
  of $b$ and $b'$ are initially $(-1,0)$ and $(1,0)$ with $b$ below
  $b'$ as in Figure~\ref{rule1}.  Let $\alpha, \beta<90^\circ$ denote
  the angles at the endpoints of $b'$. We attach the coordinate system
  to $b'$ and denote denote by $(-1+x,y)$ and $(1+u,v)$ the (moving)
  endpoints of $b$. 
  Then we have
\begin{equation}
  \label{eq:a2}
y\le 0, \
v\le 0, \
y\ge -x\tan \alpha,\
v\ge u\tan \beta,
\end{equation}
and
\begin{equation}
  \label{eq:a1}
((x-1)-(1+u))^2+ (y-v)^2=4  .
\end{equation}
{From}~\eqref{eq:a2},
we conclude that $x\ge0$, $u\le 0$, and hence $x-u\ge0$,
and furthermore
\begin{equation}
  \label{eq:a3}
  0\le -y-v \le d(x-u),
\end{equation}
with $d=\max\{\tan\alpha,\tan\beta\}\ge0$.
{From}~\eqref{eq:a1},
we obtain
\begin{align}
\label{a4}
  (x-u)^2-4(x-u)& = -(y-v)^2 \ge -(y-v)^2 -4yv
= -(-y-v)^2 \ge  -d^2(x-u)^2.
\end{align}
The last inequality is based on~\eqref{eq:a3}.
Now if $x-u\ne 0$, we can divide by $x-u$, knowing from~\eqref{eq:a3}
that $x-u> 0$, and obtain $x-u\ge 4/(1+d^2)$. Since $x$ and $u$ have to move
continuously from their starting values $x=u=0$, this is impossible. We conclude that $x-u=0$, and hence
$x=u=0$. Substituting this into~\eqref{a4}, we see that
 $-(-y-v)^2=0$ is sandwiched between two expressions which are 0.
Therefore $-y-v=0$,
 and hence
$y=v=0$.
\end{proof}

\begin{figure}
  \centering
  \begin{overpic}[scale=0.5]{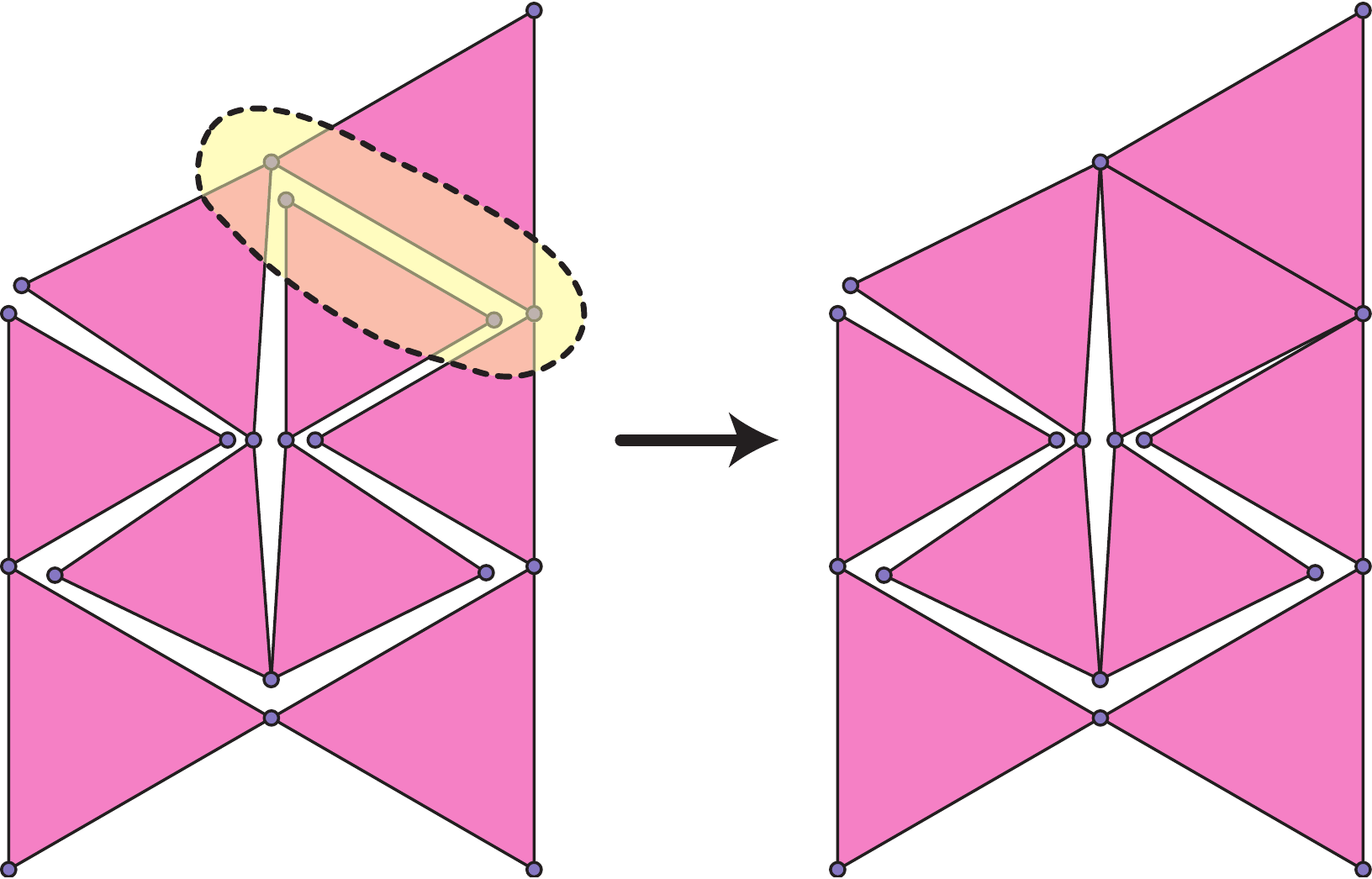}
    \put(17,58){\makebox(0,0)[b]{Rule \protect\ref{rule 1}}}
  \end{overpic}
  \caption{Applying Rule~\protect\ref{rule 1} to the chain of nine
    equilateral triangles from Figure~\protect\ref{locked 9 equilateral}.}
  \label{apply rule 1}
\end{figure}

We can apply this rule to the region shown in Figure \ref{apply rule 1},
resulting in a simpler linkage with the same infinitesimal behavior.
Although the figure shows positive separations for visual clarity,
we are in fact acting on the self-touching configuration
of Figure~\ref{locked 9 equilateral tight}.

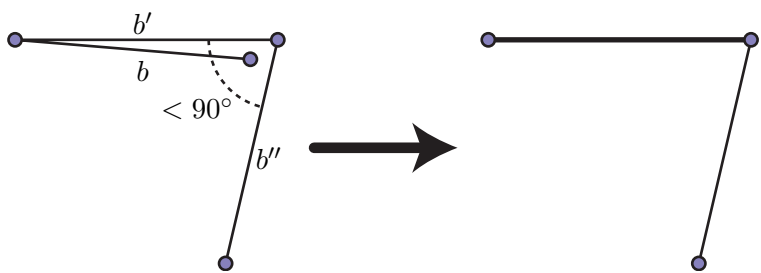
\begin{figure}
  \centering
  \begin{overpic}[scale=1]{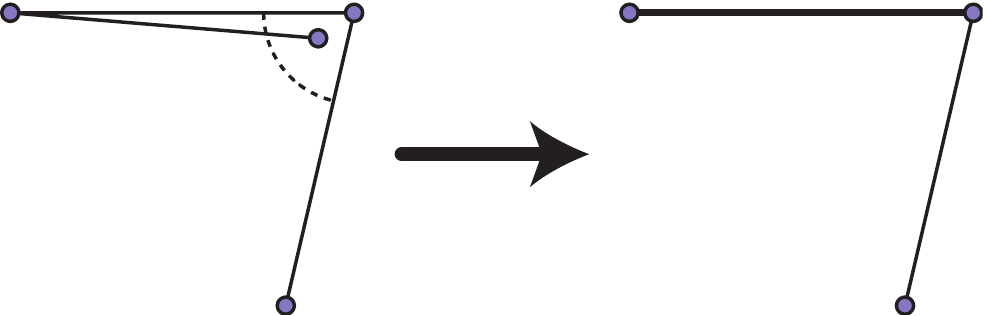}
    \put(18,31.5){\makebox(0,0)[b]{$b'$}}
    \put(18,28){\makebox(0,0)[t]{$b$}}
    \put(33,15){\makebox(0,0)[l]{$b''$}}
    \put(30,23){\makebox(0,0)[rt]{$< 90^\circ$}}
  \end{overpic}
  \caption{Rule~\protect\ref{rule 2}
           for simplifying self-touching configurations.}
  \label{rule2}
\end{figure}

\begin{onerule} \label{rule 2}
  If a bar $b$ is collocated with an incident bar $b'$ of the same length
  whose other incident bar $b''$ forms a convex angle with $b'$
  surrounding~$b$,
  then any motion must keep $b$ collocated with $b'$ for some positive time.
  See Figure~\ref{rule2}.
\end{onerule}

\begin{proof}
  The noncrossing constraints at the endpoint of $b$ surrounded by
  the convex angle formed by $b'$ and $b''$ prevent $b$ from moving relative
  to $b'$ until the convex angle opens to $\geq 90^\circ$,
  which can only happen after a positive amount of time.
The formal proof is similar (and simpler) as for Rule~\ref{rule 1}.
\end{proof}

\begin{figure}
  \centering
  \begin{overpic}[scale=0.5]{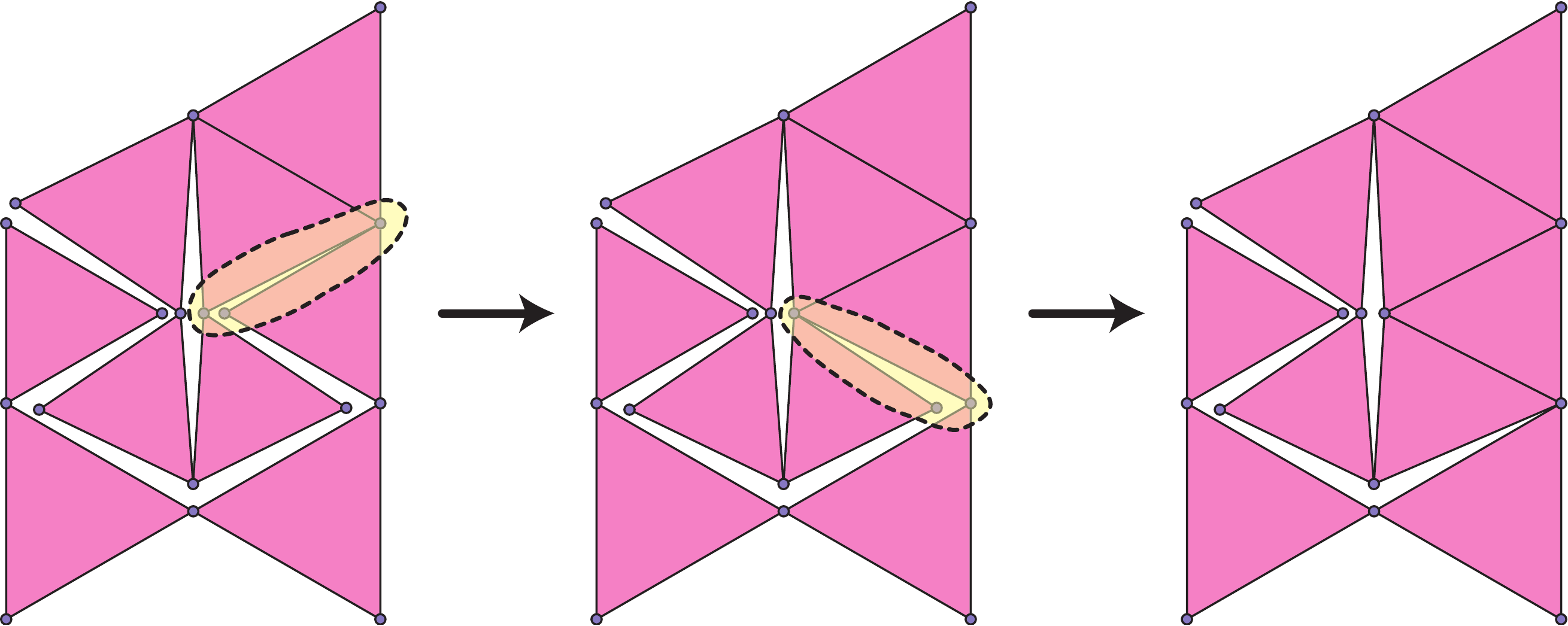}
    \put(27,26){\makebox(0,0)[l]{Rule \protect\ref{rule 2}}}
    \put(64,14){\makebox(0,0)[l]{Rule \protect\ref{rule 2}}}
  \end{overpic}
  \caption{Applying Rule~\protect\ref{rule 2} twice to the configuration
    from Figure~\protect\ref{apply rule 1}.}
  \label{apply rule 2}
\end{figure}

We can apply this rule twice, as shown in Figure \ref{apply rule 2},
to further simplify the linkage.

\begin{figure}[htbp]
  \centering
  \begin{overpic}[scale=0.5]{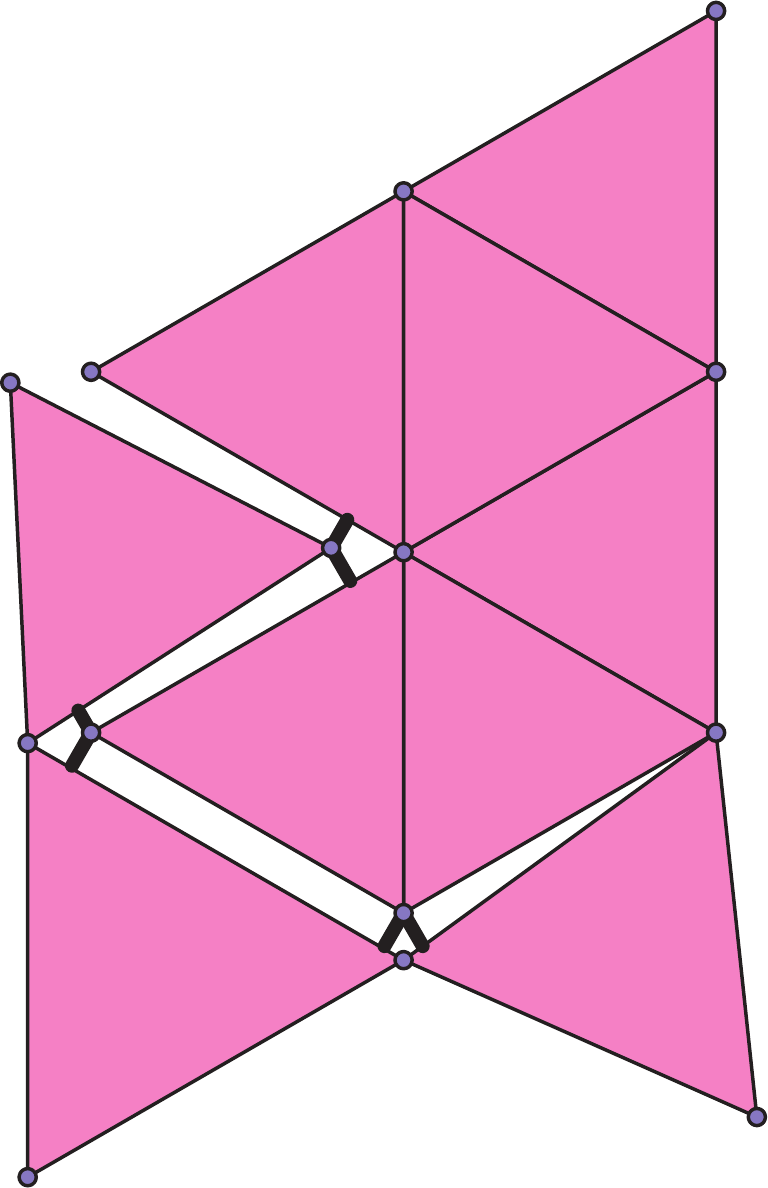}
    \put(62,38){\makebox(0,0)[l]{$A$}}
    \put(34,25){\makebox(0,0)[lb]{$B$}}
    \put(34,17){\makebox(0,0)[t]{$B'$}}
    \put(11,38.5){\makebox(0,0)[l]{$C$}}
    \put(1,37){\makebox(0,0)[r]{$C'$}}
    \put(37,54){\makebox(0,0)[l]{$D$}}
    \put(25,53.5){\makebox(0,0)[r]{$D'$}}
    \put(7.5,70){\makebox(0,0)[b]{$E$}}
  \end{overpic}
  \caption{The simplified configuration
    from Figure~\protect\ref{apply rule 2}.}
  \label{simplified}
\end{figure}
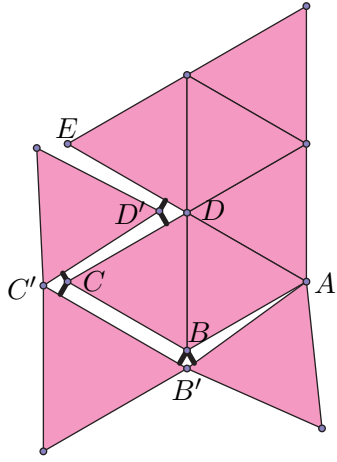

The final simplification comes from realizing that the central quadrangle gap
between triangles is effectively a triangle because the right pair of edges
are a rigid unit.  Thus the gap forms a rigid linkage (though it is not
infinitesimally rigid, because a horizontal movement of the central vertex
would maintain distances 
to the first order), so we can treat it as part of a large rigid block.
Figure~\ref{simplified} shows a simplified drawing of this self-touching
configuration, which is rigid if and only if the original self-touching
configuration is rigid.

\subsection{Stress Argument}
\label{Stress Argument}

Finally we argue that the simplified configuration of Figure~\ref{simplified}
is infinitesimally rigid using Lemma~\ref{stress lemma}.
The configuration is clearly infinitesimally rigid if
$B$~is pinned against~$B'$, $C$~is pinned against~$C'$, and
$D$~is pinned against~$D'$.
It remains to construct a stress that is negative on
all length-zero connections.
The stress we construct is nonzero only on the edges connecting points
with labels in Figure~\ref{simplified}; we also set $\omega_{A D} = 0$.

We start by assigning the stresses incident to~$A$.
We choose $\omega_{A B} < 0$ arbitrarily,
and set $\omega_{A B'} := -\omega_{A B} > 0$.
$A$ is now in equilibrium because these stress directions are parallel.

We symmetrically assign
$\omega_{B C} := \omega_{A B} < 0$ and
$\omega_{B' C'} = \omega_{A' B'} > 0$.
The resulting forces on $B$ and $B'$ are vertical.
They can be balanced by an appropriate choice of the stresses
 $\omega_{B, B' A}=\omega_{B, B' C'}<0$, which, taken together,
also point in the vertical direction.

\begin{sloppypar}
Vertex $D'$ has exactly three incident stresses---$\omega_{C' D'}$,
$\omega_{D', D C}$, and $\omega_{D', D E}$---which do not lie in a halfplane.
Thus there is an equilibrium assignment to these stresses,
unique up to scaling, and the stresses all have the same sign.
Because zero-length connections must be negative,
we are forced to make all three of these stresses negative.
We also choose this scale factor to be substantially smaller than the
stresses that have been assigned so far. 
\end{sloppypar}

By assigning $\omega_{C D} = -\omega_{C' D'}$, we establish equilibrium at
vertex $D$ as well: the forces at $D$ are the same as at $D'$, only with
reversed signs.

Vertex $C$ feels two stresses assigned so far---$\omega_{C D} > 0$ and
$\omega_{B C} < 0$.  By the choice of scale factors, the latter force
dominates, leaving us with a negative force in the direction close to $CB$,
and two stresses 
 $\omega_{C, C' B'}$ and $\omega_{C, C' D'}$ which can be used to balance this
force.
The three directions do not lie in a halfplane.
Therefore $\omega_{C, C' B'}$ and $\omega_{C, C' D'}$ can be assigned
negative stresses.

Finally, vertex $C'$ is also in equilibrium because
$\omega_{B' C'} = -\omega_{B C}$, $\omega_{C' D'} = -\omega_{C D}$,
and the stress from the zero-length connections are the same as for $C$
but in the opposite direction.

In summary, we have shown the existence of a stress that is positive on all
zero-length connections.
By Lemma~\ref{stress lemma}, the self-touching configuration is
infinitesimally rigid, so by Lemma~\ref{rigid rigid}, the configuration
is rigid.  By the simplification arguments above,
the original self-touching configuration is also rigid.
By Theorem~\ref{strong lock}, the original self-touching configuration
is strongly locked, so sufficiently perturbations are locked.

We remark that an argument similar to the one above,
using an assignment of stresses,
can also be used for proving Rules~\ref{rule 1} and~\ref{rule 2},
with an appropriate modification of Lemma~\ref{stress lemma};
however, the direct argument that we have given is simpler.

The argument relied on the isosceles triangles having an apex angle of
$< 90^\circ$ (but no more) in order to guarantee that particular triples of
stress directions are or are not in a halfplane.  It also relies on the
symmetry of the configuration through a vertical line (excluding the triangle
in the upper right).  Thus the argument generalizes to all isosceles triangles
sharper than~$90^\circ$.

\subsection{Locked Equilateral Triangles}

\begin{figure}
  \centering
  \subfigure[]
    {\includegraphics[scale=0.6]{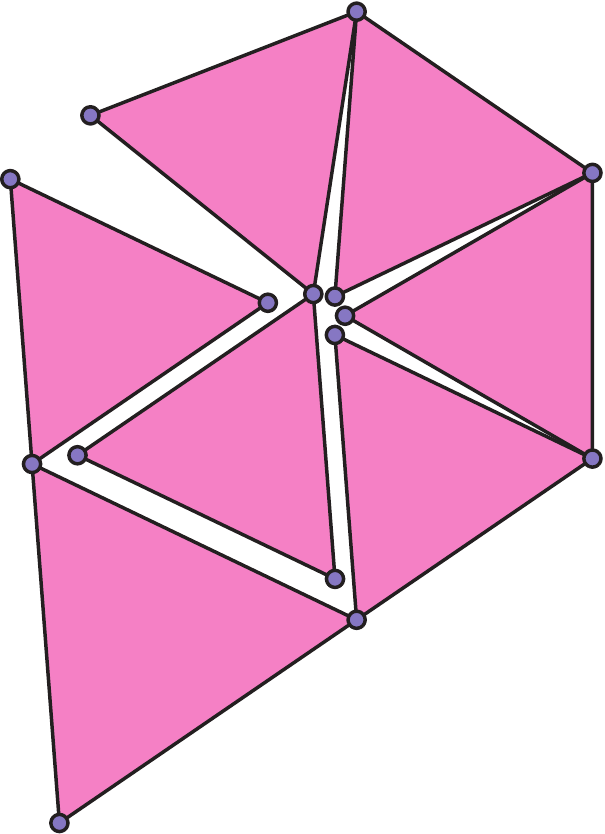}
     \label{locked 7 equilateral loose}}
  \hfil\hfil
  \subfigure[]
    {\includegraphics[scale=0.6]{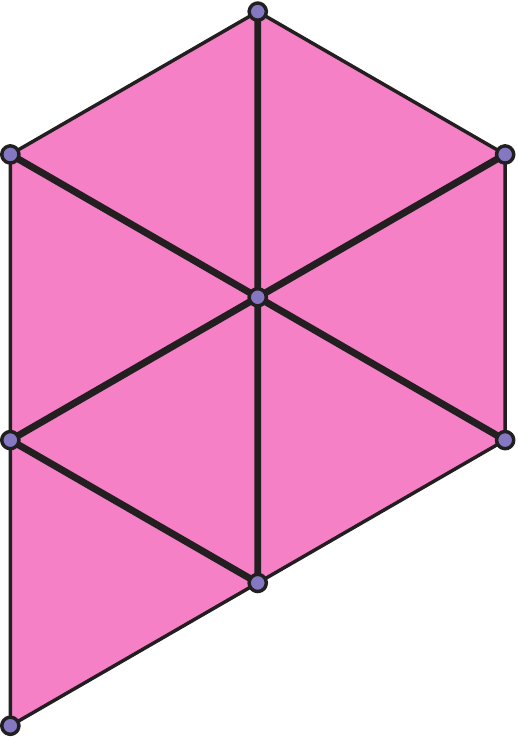}
     \label{locked 7 equilateral tight}}
  \caption{A locked chain of seven equilateral triangles.
    (a) Drawn loosely.  Separations should be smaller than they appear.
    (b) Drawn tightly, with no separation, as a self-touching configuration.}
  \label{locked 7 equilateral}
\end{figure}

Figure~\ref{locked 7 equilateral} shows another, simpler example of a locked
chain of equilateral triangles, using just seven triangles instead of nine.
However, this example cannot be stretched into a locked chain of triangles
with an arbitrary apex angle of $< 90^\circ$,
as in Figure \ref{locked 9 isosceles tight}.

\begin{figure}
  \centering
  \begin{overpic}[scale=0.5]{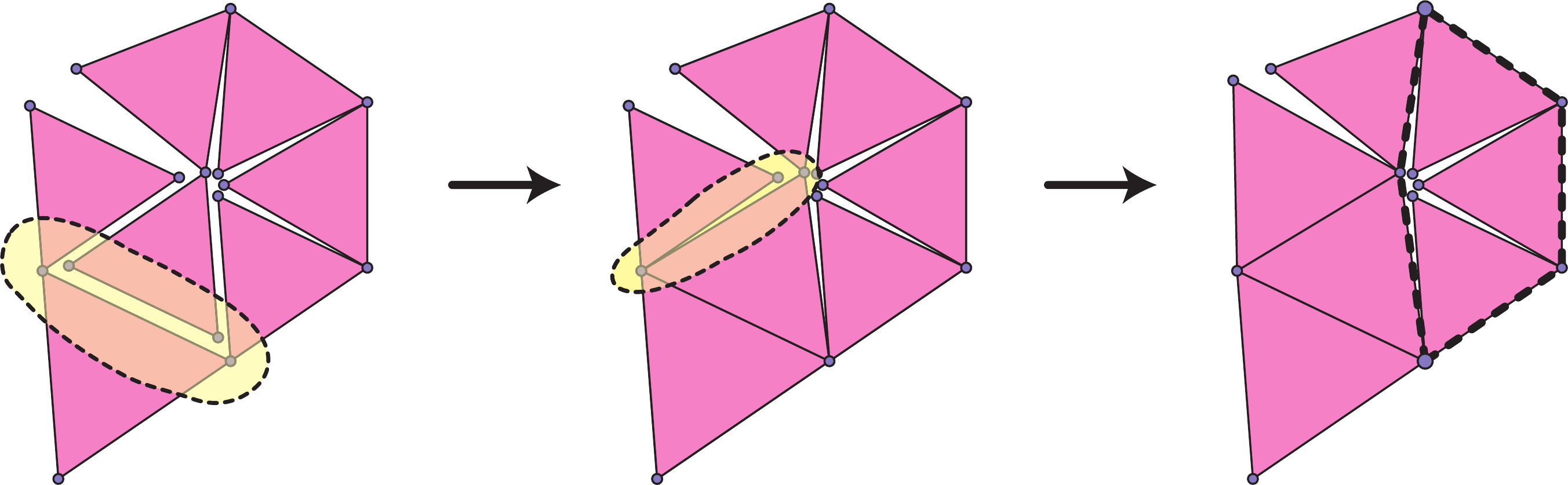}
    \put(19,8){\makebox(0,0)[l]{Rule \protect\ref{rule 1}}}
    \put(63,16){\makebox(0,0)[l]{Rule \protect\ref{rule 2}}}
    \put(91,5.5){\makebox(0,0){$A$}}
    \put(91,33){\makebox(0,0){$B$}}
  \end{overpic}
  \caption{Applying Rules~\protect\ref{rule 1} and~\protect\ref{rule 2}
    to the chain of seven equilateral triangles from
    Figure~\protect\ref{locked 7 equilateral}.}
  \label{apply rules}
\end{figure}
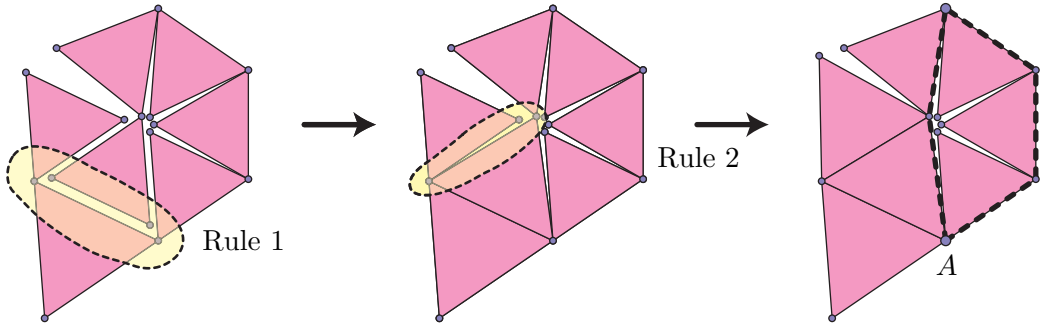

To prove that this example is locked, we first apply Rule~\ref{rule 1}
and then Rule~\ref{rule 2}, as shown in Figure~\ref{apply rules}.
Unlike the previous example, the resulting simplified configuration is not
infinitesimally rigid (the middle vertex can move infinitesimally
horizontally), so we cannot use a stress argument.
In this case, however, we can use a more direct argument to prove rigidity
of the simplified configuration (and thus of the original self-touching
configuration).

Let $\ell$ denote the side length of the triangles in any of the
self-touching configurations.
Consider the two dashed chains connecting vertices
$A$ and~$B$ in the simplified configuration.
The left chain of two bars forces the distance between
$A$ and $B$ to be at most $2 \ell$, with equality as in the original
configuration only if the angle between the two bars remains straight.
The right chain of three bars can only open its angles,
because of the three triangles on the inside,
so the right chain acts as a \emph{Cauchy arm}.
The Cauchy-Steinitz Arm Lemma (see, e.g., \cite{Connelly-1982} or \cite{Schoenberg-Zaremba-1967})
proves that the endpoints of such a chain can only get farther
away from each other.
Thus the distance between $A$ and $B$ is at least $2 \ell$,
with equality only if the angles in the right chain do not change.
These upper and lower bounds of $2 \ell$ on the distance between $A$ and $B$
force the bounds to hold with equality,
which prevents any angles from changing except possibly
for the angles at $A$ and~$B$.  However, it is impossible to change
fewer than four angles of a closed chain such as the one formed by
the left and right dashed chains.
(This simple fact was also proved by Cauchy \cite{Cromwell-1997-Cauchy}.)
Therefore, the configuration is rigid.

Applying Theorem~\ref{strong lock}, we obtain that the self-touching
configuration is strongly locked:

\begin{theorem} \label{7 triangles rigid}
  The self-touching chain of seven equilateral triangles shown in
  Figure \ref{locked 7 equilateral tight} is rigid and thus strongly locked.
  Therefore, any sufficiently small non-self-touching perturbation,
  similar to the one shown in Figure \ref{locked 7 equilateral loose},
  is locked.
\end{theorem}

\section{Conclusion}
\label{sec:conclusion}

In the time since we first reported some of our results in the form of
an extended abstract~\cite{conf-version}, a number of further
extensions have been explored. Abbott et
al.~\cite{Abbott-Abel-Charlton-Demaine-Demaine-Kominers-2008} have
utilized our results in proving that hinged dissections exist.
Abbott, Demaine and Gassend~\cite{adg-gcrts-09} provided a
generalization for the restricted case of open chains of strictly
slender adornments: even when self-touching configurations are
allowed, every chain can be opened.

A variety of open questions and extensions remain to be studied.  It
may be interesting to consider the algorithmic question of computing,
for a given configuration of an adorned chain (whose adornments are
not slender), whether or not there exists a motion that opens it.  

Our results have application to hinged dissections of polyregulars,
e.g., polyominoes; this includes the (slender) case of squares
connected at opposite corners.  An interesting question arising in
this context is whether every hinged dissection can be subdivided so
that the pieces are slender.

\begin{figure} [here]
\begin{center}
  \subfigure[]
    {\includegraphics[scale=0.8]{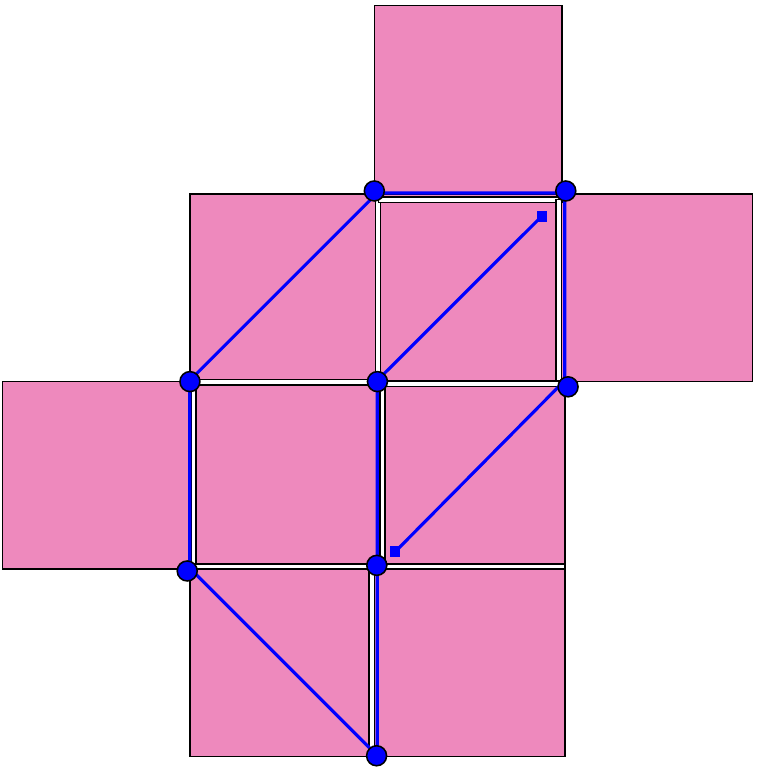}
     \label{open-locked}}
  \hfil\hfil
  \subfigure[]
    {\includegraphics[scale=0.8]{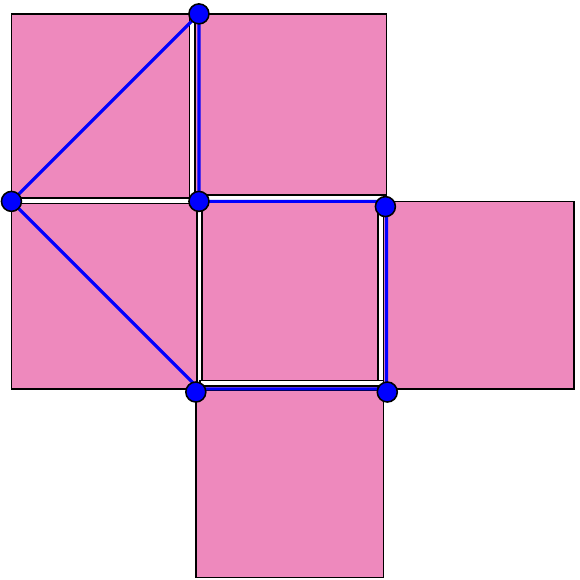}
     \label{closed-locked}}
\caption{Chains of squares that appear to be locked: (a) an open chain, and (b) a closed chain.  The squares are
self-touching; the drawing includes gaps only for clarity of illustration.}
 \label{fig:locked-squares}
\end{center}
\end{figure}

We note that there are examples of open and closed chains of
self-touching squares (some of which are attached at adjacent corners)
that we conjecture to be locked (Figure~\ref{fig:locked-squares});
while we have been unable so far to prove that they are, the
methods we employed for the locked chains of sharp triangles
(Section~\ref{locked sharp}) may be applicable.

\begin{figure} [here]
\begin{center}
\includegraphics{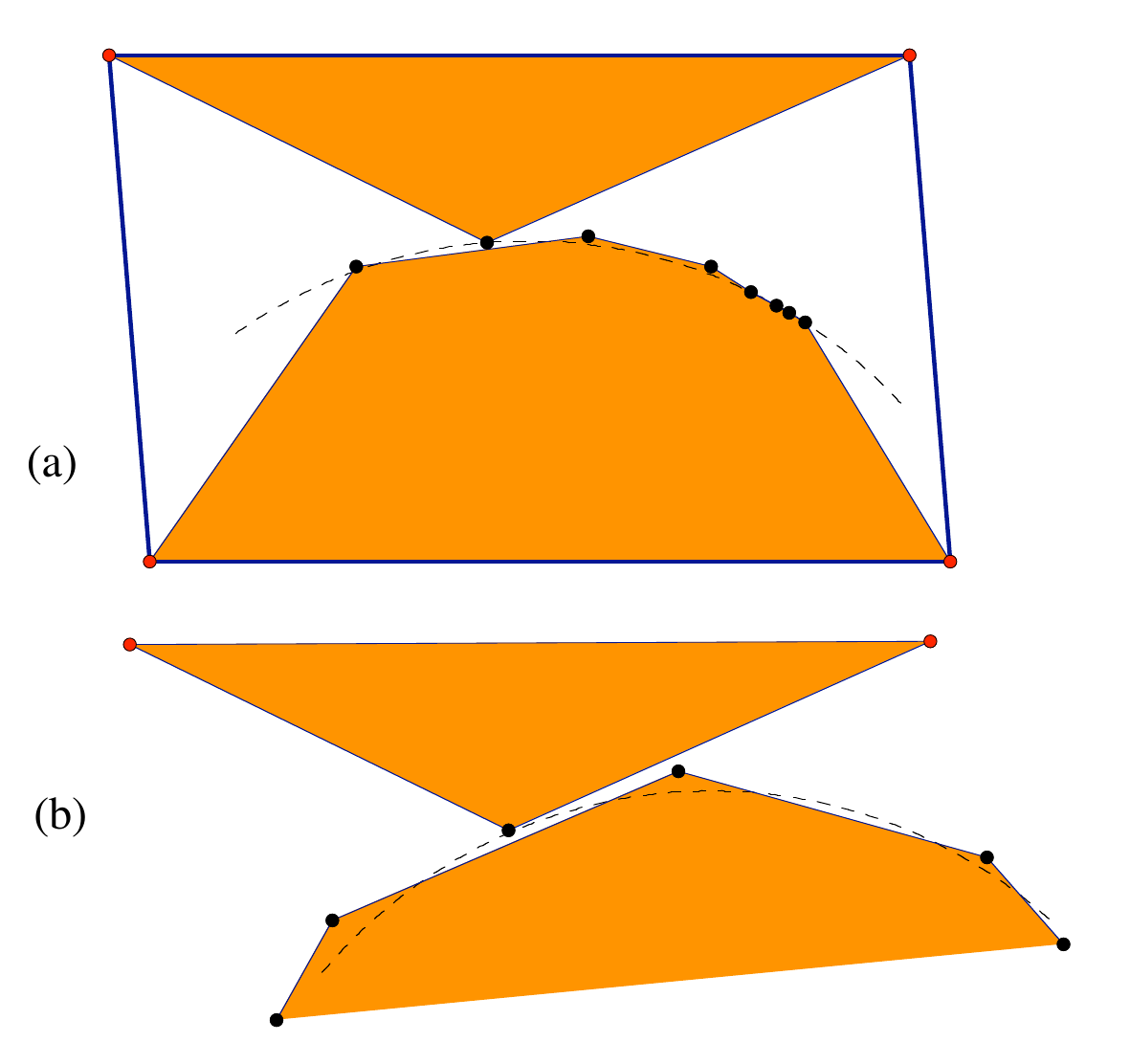}
\caption{Figure (a) shows the overall set-up of a parallelogram with two convex slender adornments attached such that the configuration space has infinitely many components.  Figure (b) is an exaggerated close-up of where the two adornments are close.}
 \label{fig:inf-comp}
\end{center}
\end{figure}

One difficulty of exploring the space of adorned
chains is highlighted by the following construction.
It consists of a closed chain, a convex parallelogram,
with slender adornments attached, where each adornment together with
its base is convex, such that the configuration space has infinitely
many components.
We attach a single obtuse triangle as a slender adornment to the top
base segment, as with Figure \ref{fig:2-components}.  If the bottom
segment is fixed the path of the bottom vertex in the upper adornment
traces out a circle, which is shown as a dashed circular arc $C$ in
Figure \ref{fig:inf-comp}(a).
The second slender adornment is attached to the bottom segment and is
the convex hull of infinitely many points, each slightly above $C$.
The points form an infinite sequence $p_1, p_2, \dots$ converging to a
point on the right $p_\infty$, and they are chosen so the straight
line interval from $p_i$ to $p_{i+1}$ intersects the lower portion of
$C$ (the open circular disk determined by $C$).  An exaggerated
picture of this construction is in Figure \ref{fig:inf-comp}(b).
Thus, the upper slender adornment intersects the lower adornment and
misses it alternately infinitely often.

\begin{thebibliography}{CvdSG{\etalchar{+}}07}

\bibitem[AAC{\etalchar{+}}08]{Abbott-Abel-Charlton-Demaine-Demaine-Kominers-20%
08}
Timothy~G. Abbott, Zachary Abel, David Charlton, Erik~D. Demaine, Martin~L.
  Demaine, and Scott~D. Kominers.
\newblock Hinged dissections exist.
\newblock In {\em Proceedings of the 24th Annual ACM Symposium on Computational
  Geometry (SoCG 2008)}, pages 110--119, College Park, Maryland, June 9--11
  2008.

\bibitem[ADG09]{adg-gcrts-09}
Timothy~G. Abbott, Erik~D. Demaine, and Blaise Gassend.
\newblock A generalized {Carpenter}'s {Rule} {Theorem} for self-touching
  linkages, January 2009.
\newblock arXiv:\href{http://arXiv.org/abs/0901.1322}{0901.1322}.

\bibitem[AKRW04]{Alt-Knauer-Rote-Whitesides-2004}
Helmut Alt, Christian Knauer, G{\"u}nter Rote, and Sue Whitesides.
\newblock On the complexity of the linkage reconfiguration problem.
\newblock In J.~Pach, editor, {\em Towards a Theory of Geometric Graphs},
  volume 342 of {\em Contemporary Mathematics}, pages 1--14. American
  Mathematical Society, 2004.

\bibitem[Ale84]{Alexander-1984}
Ralph Alexander.
\newblock The circumdisk and its relation to a theorem of {Kirszbraun} and
  {Valentine}.
\newblock {\em Math. Mag.}, 57(3):165--169, 1984.

\bibitem[AN98]{Akiyama-Nakamura-1998}
Jin Akiyama and Gisaku Nakamura.
\newblock Dudeney dissection of polygons.
\newblock In {\em Revised Papers from the Japan Conference on Discrete and
  Computational Geometry}, volume 1763 of {\em Lecture Notes in Computer
  Science}, pages 14--29, Tokyo, Japan, December 1998.

\bibitem[BC02]{Bezdek-Connelly-2002}
K\'aroly Bezdek and Robert Connelly.
\newblock Pushing disks apart---the {Kneser-Poulsen} conjecture in the plane.
\newblock {\em J. reine angew. Math.}, 553:221--236, 2002.

\bibitem[CDD{\etalchar{+}}06]{conf-version}
Robert Connelly, Erik~D. Demaine, Martin~L. Demaine, S\'andor~P. Fekete, Stefan
  Langerman, Joseph S.~B. Mitchell, Ares Rib\'o, and G{\"u}nter Rote.
\newblock Locked and unlocked chains of planar shapes.
\newblock In {\em Proceedings of the 22nd Annual Symposium on Computational
  Geometry, Sedona}, pages 61--70. Association for Computing Machinery, 2006.
\newblock arXiv:\href{http://arXiv.org/abs/cs/0604022}{cs/0604022}.

\bibitem[CDR02]{Connelly-Demaine-Rote-2002-infinitesimally-locked}
Robert Connelly, Erik~D. Demaine, and G\"unter Rote.
\newblock Infinitesimally locked self-touching linkages with applications to
  locked trees.
\newblock In J.~Calvo, K.~Millett, and E.~Rawdon, editors, {\em Physical Knots:
  Knotting, Linking, and Folding of Geometric Objects in 3-space}, pages
  287--311. American Mathematical Society, 2002.

\bibitem[CDR03]{Connelly-Demaine-Rote-2003}
Robert Connelly, Erik~D. Demaine, and G\"unter Rote.
\newblock Straightening polygonal arcs and convexifying polygonal cycles.
\newblock {\em Discrete \& Computational Geometry}, 30(2):205--239, September
  2003.

\bibitem[Con82]{Connelly-1982}
Robert Connelly.
\newblock Rigidity and energy.
\newblock {\em Invent. Math.}, 66(1):11--33, 1982.

\bibitem[Con08]{Bobs-survey}
Robert Connelly.
\newblock Expansive motions.
\newblock In Jacob~E. Goodman, J\'anos Pach, and Richard Pollack, editors, {\em
  Surveys on Discrete and Computational Geometry---Twenty Years Later}, volume
  453 of {\em Contemporary Mathematics}, pages 213--229. American Mathematical
  Society, 2008.

\bibitem[Cro97]{Cromwell-1997-Cauchy}
Peter~R. Cromwell.
\newblock Equality, rigidity, and flexibility.
\newblock In {\em Polyhedra}, chapter~6, pages 219--247. Cambridge University
  Press, 1997.

\bibitem[Csi01]{Csikos-2001}
Bal\'azs Csik\'os.
\newblock On the volume of flowers in space forms.
\newblock {\em Geom. Dedicata}, 86(1-3):59--79, 2001.

\bibitem[CvdSG{\etalchar{+}}07]{csgor-ihp-06}
Jae-Sook Cheong, A.~Frank van~der Stappen, Ken Goldberg, Mark~H. Overmars, and
  Elon Rimon.
\newblock Immobilizing hinged polygons.
\newblock {\em International Journal on Computational Geometry and
  Applications}, 17(1):45--70, 2007.

\bibitem[DDE{\etalchar{+}}05]{Demaine-Demaine-Eppstein-Frederickson-Friedman-2%
005}
Erik~D. Demaine, Martin~L. Demaine, David Eppstein, Greg~N. Frederickson, and
  Erich Friedman.
\newblock Hinged dissection of polyominoes and polyforms.
\newblock {\em Computational Geometry: Theory and Applications},
  31(3):237--262, June 2005.

\bibitem[DDLS05]{Demaine-Demaine-Lindy-Souvaine-2005}
Erik~D. Demaine, Martin~L. Demaine, Jeffrey~F. Lindy, and Diane~L. Souvaine.
\newblock Hinged dissection of polypolyhedra.
\newblock In {\em Proceedings of the 9th Workshop on Algorithms and Data
  Structures}, volume 3608 of {\em Lecture Notes in Computer Science}, pages
  205--217, Waterloo, Canada, August 2005.

\bibitem[Dud02]{Dudeney-1902-hinged}
Henry~E. Dudeney.
\newblock Puzzles and prizes.
\newblock {\em Weekly Dispatch}, 1902.
\newblock The puzzle appeared in the April 6 issue of this column. An unusual
  discussion followed on April 20, and the solution appeared on May 4.

\bibitem[Epp01]{Eppstein-2001-mirror-dissection}
David Eppstein.
\newblock Hinged kite mirror dissection, June 2001.
\newblock arXiv:\href{http://arXiv.org/abs/cs.CG/0106032}{cs.CG/0106032}.

\bibitem[Fre02]{Frederickson-2002}
Greg~N. Frederickson.
\newblock {\em Hinged Dissections: Swinging \& Twisting}.
\newblock Cambridge University Press, August 2002.

\bibitem[GM95]{Gordon-Meyer-1995}
Y.~Gordon and M.~Meyer.
\newblock On the volume of unions and intersections of balls in {Euclidean}
  space.
\newblock In Joram Lindenstrauss and Vitali Milman, editors, {\em Geometric
  Aspects of Functional Analysis. Israel Seminar (GAFA) 1992--94}, volume~77 of
  {\em Operator Theory: Advances and Applications}, pages 91--101.
  Birkh\"auser, 1995.

\bibitem[Kir34]{Kirszbraun-1934}
M.~D. Kirszbraun.
\newblock {\"Uber} die zusammenziehende und {Lipschitzsche} {Transformationen}.
\newblock {\em Fundamenta Mathematicae}, 22:77--108, 1934.

\bibitem[MTW{\etalchar{+}}02]{Mao-Thallidi-Wolfe-Whitesides-Whitesides-2002}
Chengde Mao, Venkat~R. Thallidi, Daniel~B. Wolfe, Sue Whitesides, and George~M.
  Whitesides.
\newblock Dissections: Self-assembled aggregates that spontaneously reconfigure
  their structures when their environment changes.
\newblock {\em Journal of the American Chemical Society}, 124:14508--14509,
  2002.

\bibitem[Rei79]{r-cmpg-79}
John~H. Reif.
\newblock Complexity of the mover's problem and generalizations.
\newblock In {\em Proceedings of the 20th Annual IEEE Symposium on Foundations
  of Computer Science}, pages 421--427, 1979.

\bibitem[Str00]{Streinu-2000}
Ileana Streinu.
\newblock A combinatorial approach to planar non-colliding robot arm motion
  planning.
\newblock In {\em Proceedings of the 41st Annual Symposium on Foundations of
  Computer Science}, pages 443--453, Redondo Beach, California, November 2000.

\bibitem[Str05]{Streinu-2005}
Ileana Streinu.
\newblock Pseudo-triangulations, rigidity and motion planning.
\newblock {\em Discrete \& Computational Geometry}, 34(4):587--635, November
  2005.

\bibitem[SZ67]{Schoenberg-Zaremba-1967}
I.~J. Schoenberg and S.~K. Zaremba.
\newblock On {Cauchy}'s lemma concerning convex polygons.
\newblock {\em Canadian Journal of Mathematics}, 19(4):1062--1071, 1967.

\end{thebibliography}
\bibliographystyle{alpha}

\newcommand{\etalchar}[1]{$^{#1}$}

\end{document}